\documentclass[aps,11pt,english,tightenlinesletterpaper,superscriptaddress,secnumarabic,nofootinbib]{revtex4}

\pdfoutput=1

\usepackage{multirow}
\usepackage{mypackages}
\usepackage[sort&compress]{natbib}
\usepackage{hyperref} 	%Automatically links \label and \ref commands; Always load last
\usepackage[all]{hypcap}  %Ensures figure hyperlinks are positioned to display entire figure
\hypersetup{
    colorlinks=true,		%false: black links; true: colored links
    linkcolor=blue,		%color of internal links
    citecolor=green,		%color of links to bibliography
    filecolor=magenta,	%color of file links
    urlcolor=cyan		%color of external links
}

\usepackage{myterms}

\newcommand{\eps}{\epsilon}
\newcommand{\tb}[1]{\tilde{\bar{#1}}}
\newcommand{\bt}[1]{\bar{\tilde{#1}}}
\newcommand{\G}{\tilde{G}}
\newcommand{\X}{\tilde{X}}
\newcommand{\BV}{{\text{\sc bv}}}
\newcommand{\LV}{{\text{\sc lv}}}
\newcommand{\CP}{{\text{\sc cp}}}
\newcommand{\LP}{{\text{\sc lp}}}
\newcommand{\MSSM}{{\text{\sc mssm}}}

\newcommand{\eBV}{\epsilon_{\BV}}

\begin{document}

\title{Gravitino Leptogenesis}

\author{Lawrence M. Krauss}
\email{krauss@asu.edu}
\affiliation{\small Department of Physics and School of Earth and Space Exploration \\ Arizona State University, Tempe, AZ 85827-1404}
\affiliation{\small Research School of Astronomy and Astrophysics, Mt. Stromlo Observatory, \\ Australian National University, Canberra, Australia 2614}

\author{Andrew J. Long}
\email{andrewjlong@asu.edu}
\affiliation{\small Department of Physics and School of Earth and Space Exploration \\ Arizona State University, Tempe, AZ 85827-1404}

\author{Subir Sabharwal}
\email{subir.sabharwal@asu.edu}
\affiliation{\small Department of Physics and School of Earth and Space Exploration \\ Arizona State University, Tempe, AZ 85827-1404}

\begin{abstract}
We explore here a new mechanism by which the out of equilibrium decay of heavy gravitinos, followed by  possible R-parity violating decays in the Minimal Supersymmetric Standard Model (MSSM) can generate the baryon asymmetry of the universe.  In this mechanism, gravitino decay produces a CP-asymmetry that is carried by squarks or sleptons.  These particles then decay through R-parity violating operators generating a lepton asymmetry.  The lepton asymmetry is converted into a baryon asymmetry by weak sphalerons, as in the familiar case of leptogenesis by Majorana neutrino decays.  To ensure that the gravitino decays while the sphaleron is still in equilibrium, we obtain a lower bound on the gravitino mass, $m_{3/2} \gtrsim 10^{8} \GeV$, and therefore our mechanism requires a high scale of SUSY breaking, as well as minimum reheating temperature after inflation of $T\gtrsim 10^{12} \GeV$ in order to for the gravitino density to be sufficiently large to generate the baryon asymmetry today.  
We consider each of the MSSM's relevant R-parity violating operators in turn, and derive constraints on parameters in order to give rise to a baryon asymmetry comparable to that observed today, consistent with low energy phenomenological bounds on SUSY models.
\end{abstract}


%\preprint{}
%\begin{document}
\maketitle

\setlength{\parindent}{20pt}
\setlength{\parskip}{2.5ex}

%===========================
% Introduction
%===========================
\section{Introduction}\label{sec:Introduction}

%===========
The gravitino has two features that, taken together, make it unique among the particles of the Minimal Supersymmetric Standard Model (MSSM): the gravitino mass is directly related to the scale of supersymmetry breaking, and its interactions are fixed to be uniform and of gravitational strength.  
Together these properties imply that if it is kinematically allowed for the gravitino to decay, then its decay will necessarily be out of equilibrium.  
That is to say, the inverse decay process occurs with a rate $\Gamma_{inv} \sim T^3 / M_P^2$, which is always smaller than the Hubble expansion rate $H \sim T^2 / M_P$.  
A departure from thermal equilibrium is one of the three necessary conditions for the creation of the baryon asymmetry of the universe (BAU) \cite{Sakharov:1967dj}.  
The two remaining conditions, the violation of CP and the violation of baryon number (B), are already present in the MSSM through SUSY-breaking and R-parity violating operators.  
This makes the gravitino a prime candidate with which to study the origin of the cosmological baryon asymmetry.  

%===========
A possible connection between the gravitino and the cosmic baryon asymmetry was first identified by Cline and Raby \cite{Cline:1990bw}, hereafter denoted as CR, who recognized that the so-called `gravitino problem' and the problem of the cosmic baryon asymmetry could have a common solution.  
If the gravitino decays prior to the onset of Big Bang Nucelosynthesis (BBN) at $T_{BBN} \approx 1 \MeV$, then the abundance of light elements is not disrupted and the gravitino problem is avoided \cite{Pagels:1981ke, Weinberg:1982zq, Ellis:1984eq, Kawasaki:1994af}.  
The gravitino decays with a rate \cite{Krauss:1983ik}
\begin{align}
	\Gamma_{3/2} = \frac{N_{\rm eff}}{2\pi} \frac{m_{3/2}^3}{M_P^2}
\end{align}
where $N_{\rm eff}$ is the effective number of decay channels, $m_{3/2}$ is the gravitino mass, and $M_P \approx 2.4 \times 10^{18} \GeV$ is the reduced Planck mass.  
Therefore, a gravitino with mass $m_{3/2} \ge O(10 \TeV)$ will safely decay at a temperature $T_d \ge T_{BBN}$.  
CR supposed that these gravitinos decay through the MSSM's B-violating operator 
\begin{align}
	W_{\text{\sc bv}} = \frac{1}{2} \lambda^{\prime \prime} \hat{U}^c \hat{D}^c \hat{D}^c \com
\end{align}
(see, e.g., \cite{Barbier:2004ez} for a review of R-parity violation), and showed that such a decay could give rise to the cosmological baryon asymmetry.  

%===========
The present work explores how a baryon asymmetry can be generated from gravitino decays at a different scale, without the aid of the MSSM's B-violating operator.  
If gravitino decay gives rise to a lepton asymmetry, this asymmetry can be transferred to the baryons by the weak sphaleron process \cite{Kuzmin:1985mm}, which violates the anomalous $B+L$ charge conservation and rapidly coverts L-number into B-number, as in the by now standard leptogenesis scenario in which the lepton asymmetry arises from the out of equilibrium decay of the Majorana neutrino \cite{Fukugita:1986hr} (see also \cite{Buchmuller:2005eh} for a review).  

%===========
In some sense, gravitino leptogenesis is more general than either the case considered by CR, or standard leptogenesis, as the gravitino can decay through one of the MSSM's three lepton number violating operators:  
\begin{align}
	W_{\text{\sc lv}} &= 
	\frac{1}{2} \lambda \hat{L} \hat{L} \hat{E}^c 
	+ \frac{1}{2} \lambda^{\prime} \hat{L} \hat{Q} \hat{D}^c
	+ \mu^{\prime} \hat{H}_u \hat{L} \per
\end{align}
Since weak sphalerons go out of equiliburm after electroweak symmetry breaking occurs at $T_{ew} \approx 100 \GeV$ \cite{Kuzmin:1985mm}, to ensure that the gravitino decays prior to this time, we obtain a lower bound on the gravitino mass: $m_{3/2} \gtrsim 10^{8} \GeV$ (see \sref{sec:CosmoContext} for details).  
Therefore, a gravitino leptogenesis mechanism can be operative for models with a high SUSY-breaking scale, $M_S \gtrsim 10^{13} \GeV$.  

%===========
The organization of this paper is as follows.  
In \sref{sec:CosmoContext} we discuss the cosmological context of gravitino leptogenesis, and specifically we derive bounds on the gravitino mass and reheat temperature which are imposed by requiring that gravitinos decay prior to electroweak symmetry breaking.  
In \sref{sec:LG_Channels} we consider each of the MSSM R-parity violating operators in turn, including a brief review of the B-violating operator that was previously studied by CR.  
For the other operators, we determine the parameter ranges required to generate the baryon asymmetry of the universe through a gravitino leptogenesis mechanism.  
In \sref{sec:Conclusions} we summarize our conclusions.  
\aref{app:Notation} summarizes the notation that we use, and \aref{app:GravitinoBaryogenesis} outlines the details of the calculational scheme presented by CR.

%===========================
% Cosmological Context of Gravitino Decays
%===========================
\section{Cosmological Context of Gravitino Decays}\label{sec:CosmoContext}

%===========
In this section, we will derive an expression for the baryon asymmetry of the universe in terms of i) the gravitino mass, ii) the reheat temperature after inflation, and iii) a parameter $\beta$ which controls the branching fraction of gravitino decays into L-number.  
In the following section we then present detailed estimates of $\beta$ for each of the MSSM's R-parity violating operators.  

%-------------------------
%  Gravitino Production
%-------------------------
\subsection{Gravitino Production}\label{sub:GravProd}

%===========
Inflation dilutes any primordial gravitino abundance, but during reheating gravitinos are regenerated by interactions in the hot plasma.  
This regeneration occurs at $T \approx T_{RH}$, which is generically much earlier than gravitino decay.  
During adiabatic expansion following inflation, the gravitino to entropy ratio $Y_{3/2} \equiv n_{3/2} / s$ is conserved.  We denote the number density of gravitinos by $n_{3/2}$ and $s = \frac{2 \pi^2}{45} g_{\ast}(T) \, T^3$ is the entropy density of the plasma, where $g_{\ast}$ is the number of helicity states with equilibrium number density in the relativistic gas at temperature $T$. At the (later) time of gravitino decay, when all of the SM species are light and all but a few of the superpartners are heavy and decoupled, we have $g_{\ast} \gtrsim 100$. By summing the various production processes and solving the thermally averaged Boltzmann equation the gravitino relic abundance has been estimated to be \cite{Ellis:1984eq, Kawasaki:1994af} 
\begin{align}\label{eq:def_Y32}
	Y_{3/2} 
	= \left( \frac{45 \zeta(3)}{2 \pi^4} \frac{1}{g_{\ast}(T_{RH})} \right)^2 \frac{s(T_{RH}) \langle \sigma_{3/2} v \rangle}{H(T_{RH})} \per
\end{align}
In making this estimate the universe is assumed to be radiation dominated with a Hubble parameter $H(T) \approx \sqrt{\pi^2 g_{\ast}(T)/90} \, T^2 / M_P$. 

The thermally averaged gravitino production cross section can be written as $\langle \sigma_{3/2} v \rangle = C_i g_i^2 M_{P}^{-2}$, where $g_i$ represent gauge couplings and the sum runs over gauge groups, and $C_i = O(1-10)$ \cite{Kawasaki:1994af}.  
This result reflects the fact that gravitinos only interact with a gravitational strength and that the higher-spin gauge fields are required to build up the spin-3/2 gravitino.  
The largest contribution comes from the $\SU{3}$ group for which $C_{3} \approx 10$ and $g_{3} \approx 1$, and therefore we will conservatively assume $\langle \sigma_{3/2} v \rangle = 10 M_{P}^{-2}$.  
This gives the gravitino relic abundance estimate
\begin{align}\label{eq:estimate_Y32}
	Y_{3/2} \approx 10^{-12} \frac{T_{RH}}{10^{10} \GeV}. 
\end{align}
In order for gravitinos to be sufficiently abundant to account for the baryon asymmetry, $Y_{B} \sim 10^{-10}$, the reheat temperature must be sufficiently high, $T_{RH} \gtrsim 10^{12} \GeV$.

%-------------------------
%  Gravitino Decay
%-------------------------
\subsection{Gravitino Decay}\label{sub:GravDecay}

%===========
As can be seen from \eref{eq:estimate_Y32}, the relic gravitino abundance is generally low, and therefore presumably the gravitinos can decay before they come to dominate the energy density of the universe.  
We can confirm this expectation by comparing the energy density of radiation, $\rho = \frac{3}{4} s T$, with the energy density in gravitinos, $\rho_{3/2} = m_{3/2} n_{3/2}$.  
Imposing $\rho_{3/2} = \rho$ gives
\begin{align}\label{eq:def_Teq}
	T_{\rm eq} = \frac{4}{3} Y_{3/2} m_{3/2}
	\approx 0.1 \MeV \frac{T_{RH}}{10^{10} \GeV}  \frac{m_{3/2}}{10^{8} \GeV}
\per
\end{align}
Since successful gravitino leptogenesis requires the gravitino to decay prior to electroweak symmetry breaking at $T_{ew} \sim 10^{2} \GeV$, it is clear in light of \eref{eq:def_Teq} that the universe will be radiation dominated at the time of gravitino decay.  

%===========
The couplings of the gravitino are only gravitational strength, and its decay rate is given by \cite{Krauss:1983ik}
\begin{align}
	\Gamma_{3/2} \approx \frac{N_{\rm eff}}{2\pi} \frac{m_{3/2}^3}{M_P^2} \per
\end{align}
where $N_{\rm eff}$ is the effective number of decay channels.  
If all the decay products are much lighter than the gravitino, then $N_{\rm eff}$ just counts the total number of kinematically allowed channels with relative weighting between the chiral superfield and vector superfield final states due to helicity:  
\begin{align}\label{eq:Neff_def}
	N_{\rm eff} = N_{V} + \frac{1}{12} N_{\chi}
\end{align}
where $N_{V}$ is the number of vector superfields and $N_{\chi}$ is the number of chiral superfields into which the gravitino can decay.  
If the entire MSSM particle content is light compared to the gravitino, then $N_{V} = 1 + 3 + 8 = 12$ and $N_{\chi} = 36 + 9 + 4 = 49$, which give $N_{\rm eff} \simeq 16$.  

We will work in an instantaneous decay approximation with a gravitino lifetime is $\tau_{3/2} = \Gamma_{3/2}^{-1}$.   
The temperature $T_d$ at which gravitino decays take place is given by
\begin{align}
	\tau = t_{\rm age}(T_d)
\end{align}
where $t_{\rm age}(T)$ is the age of the universe as a function of temperature.  
During the radiation dominated era
\begin{align}
	t_{\rm age}(T) =
	\frac{1}{2} \frac{1}{H} = \sqrt{ \frac{45}{2\pi^2 g_{\ast}} } \frac{M_P}{T^2} 
\end{align}
where we have used $H = \sqrt{\rho/3M_P^2}$.  
Solving for $T_d$ one finds
\begin{align}\label{eq:Tdecay}
	T_d = 
	m_{3/2} \left( \frac{3}{2} \frac{N_{\rm eff}}{\pi^2} \sqrt{ \frac{5}{2 g_{\ast}} } \frac{m_{3/2}}{M_P} \right)^{1/2} 
	\simeq 400 \GeV \left( \frac{m_{3/2}}{10^{8} \GeV} \right)^{3/2}.
\end{align}

%===========
Let us now discuss the constraints on $m_{3/2}$ and $T_{RH}$.  
Typically these parameters are constrained by the requirement that stable gravitinos do not overclose the universe  \cite{Pagels:1981ke, Weinberg:1982zq} or the requirement that late decaying gravitinos do not disrupt the abundance of light elements or distort the cosmic microwave background \cite{Ellis:1984eq, Kawasaki:1994af}.  
In our model, the gravitino must satisfy an even more stringent requirement.  It must decay before the electroweak phase transition takes place when the electroweak sphalerons are still in equilibrium.  
Imposing $T_d \gtrsim 100 \GeV$, we then obtain a lower bound on the gravitino mass 
\begin{align}\label{eq:m32bound}
	m_{3/2} \gtrsim 10^8 \GeV \per
\end{align}
It is evident that successful gravitino leptogenesis requires an especially heavy gravitino.  
Since the gravitino mass is set directly by the scale of SUSY breaking, $M_S$, through the relationship
\begin{align}
	m_{3/2} \approx \frac{M_S^2}{M_P} \com
\end{align}
the bound \eref{eq:m32bound} implies 
\begin{align}\label{eq:MSbound}
	M_S \gtrsim 10^{13} \GeV \per
\end{align}
Such a large scale of SUSY-breaking is less theoretically attractive than connecting it to the weak scale in order to resolve the hierarchy problem, but not only is such a scenario is not ruled out on empirical grounds \cite{Pagels:1981ke, Weinberg:1982zq}, weak scale SUSY breaking models are becoming more tightly constrained due to the absence of SUSY-induced effects at colliders, including the LHC.  High scale SUSY breaking models have in fact already been considered  as interesting alternatives to weak scale SUSY for other reasons\cite{ArkaniHamed:2004fb, Giudice:2004tc, Arvanitaki:2012ps}.  

%===========
There is no direct empirical probe of the reheat temperature, and the only hard bound is $T_{RH} > T_{BBN} \approx 1 \MeV$.  
However, energy conservation arguments relate the reheat temperature to the energy scale of inflation $V_{\rm inf} = 3 M_P^2 H_{\rm inf}^2 \sim T_{RH}^4$.  
The scale of inflation, in turn, is probed by tensor perturbations in the cosmic microwave background (CMB) radiation power spectrum.  
In particular, the ratio of the amplitudes of tensor and scalar perturbations, $r = \Delta_h^2 / \Delta_{\mathcal{R}}^2$, is, for single field inflation models,  generally proportional to the energy scale of inflation $V_{\rm inf} \approx (10^{16} \GeV)^4 (r/.01)$ \cite{Krauss:1992ke} \cite{Liddle:1993ch} \cite{Krauss:2010ty}. Precisions measurements of the CMB by the Planck satellite yield the bound $r < 0.11$ at 95 \% CL \cite{Ade:2013rta}.  
This translates into an upper bound on the reheat temperature,
\begin{align}\label{eq:TRHbound}
	T_{RH} \lesssim O(10^{16} \GeV) \com
\end{align}
which in turn implies an upper bound on the gravitino relic abundance, $Y_{3/2} \lesssim O(10^{-6})$, via \eref{eq:estimate_Y32}.  

%-------------------------
%  Lepton Number Generation
%-------------------------
\subsection{Lepton Number Generation}\label{sub:LNumberGeneration}

%===========
The gravitino decay rate, described earlier, is estimated through its dominant, standard decay channels.  
We now suppose that the gravitino has a number of subdominant decay channels which are mediated by CP- and L-number-violating interactions.   This will provide the conditions necessary for the creation of a lepton asymmetry $n_{L}$, which will be proportional to the number density of gravitinos before they decay multiplied by the branching ratio for the decays which lead to the asymmetry.  
We can therefore write: 
\begin{align}\label{eq:def_beta}
	n_{L}(T_{a}) = \beta \, n_{3/2}(T_{d})
\end{align}
where $T_a > T_d$ is the temperature after gravitino decays (see below), and the parameter $\beta$ is defined as the weighted branching fraction
\begin{align}\label{eq:def_beta_from_BF}
	\beta = \sum_{i} L \left[ \left\{ f_i \right\} \right] {\rm BR}\left( \tilde{G} \to \left\{ f_i \right\} \right)
\end{align}
where the sum runs over all possible final states labeled by $\left\{ f_i \right\}$, and $L[\left\{ f_i \right\}]$ is the lepton number of a given final state (possibly negative).  
In \sref{sec:LG_Channels} we will evaluate \eref{eq:def_beta_from_BF} in terms of the model parameters. 
For now, we will simply treat $\beta$ as a free parameter.  

%===========
Since presumably the L-violating decays are rare, we expect $\beta \ll 1$.  
In the instantaneous decay approximation, the energy density is conserved, and 
all of the energy in the gravitinos is transferred to radiation.   Since the number of thermalized species remains unchanged (since $M_{SUSY} \gg T_{d} \gg M_{EW}$), the plasma heats up as a result of the energy injection \cite{Scherrer:1984fd}.  
The energy density after decay $\rho$ is given by 
\begin{align}\label{eq:rho_after}
	\rho(T_{a}) = \rho^{(0)}(T_d) + \rho_{3/2}(T_d) 
\end{align}
where $T_a$ is the temperature after the gravitinos decay and we use the superscript $``(0)"$ to denote the era prior to gravitino decay.  
The lepton asymmetry is parametrized by 
\begin{align}
	Y_{L} = \frac{n_L}{s} \per
\end{align}
Using Eqns.~(\ref{eq:def_beta}) and (\ref{eq:rho_after}) and $s = (4/3) \rho/T$, we can evaluate the lepton asymmetry as
\begin{align}
	Y_{L} = \frac{\beta \, Y_{3/2}}{\left( 1 + \frac{4}{3} \frac{m_{3/2}}{T_d} Y_{3/2} \right)^{3/4}} \per
\end{align}
Using Eqns.~(\ref{eq:estimate_Y32}) and (\ref{eq:Tdecay}), we see that $Y_L$ is a function of the free parameters $m_{3/2}$, $T_{RH}$, and $\beta$.  
For the allowed range of parameters, $m_{3/2} \gtrsim 10^{8} \GeV$ and $T_{RH} \lesssim 10^{16} \GeV$, the ratio $m_{3/2} Y_{3/2} / T_d \ll 1$ and the entropy injection is negligible.  
Also, provided that $T_a \gg T_{EW}$, the EW sphalerons will then efficiently convert L-number into B-number.  
The final baryon asymmetry is obtained from $Y_L$ after multiplying by a factor of $-8/23$ \cite{Khlebnikov:1988sr} to obtain
\begin{align}\label{eq:YB_estimate}
	Y_{B} \approx -\frac{8}{23} \beta Y_{3/2}
\per
\end{align}
Note that $Y_B > 0$ requires $\beta < 0$.  

%==========
\begin{figure}[t]
\vspace{0.2in}
\begin{center}
\includegraphics[width=0.5\textwidth]{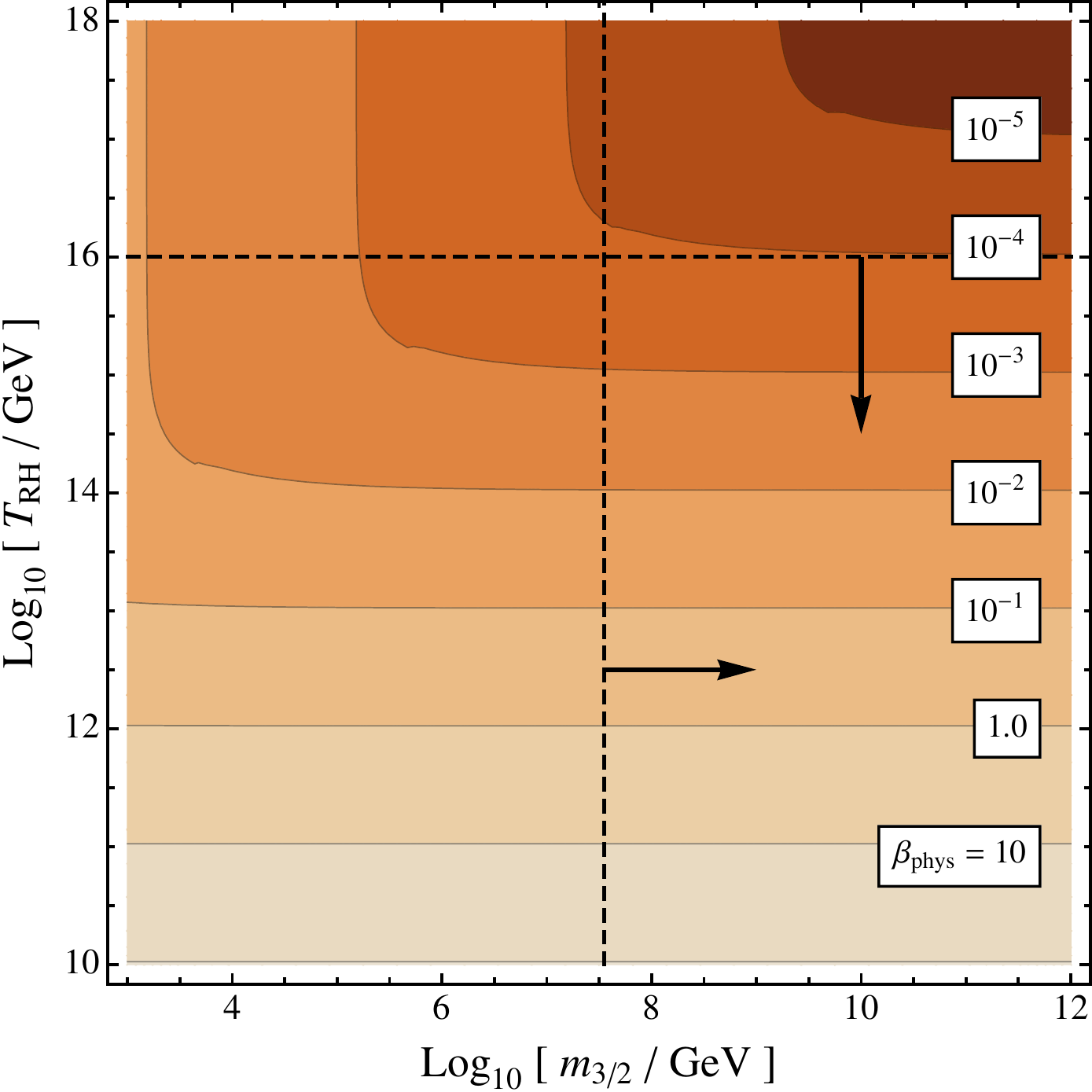} 
\caption{
\label{fig:betaphys}
The value of $\beta$ required to ensure $Y_{B} = (Y_{B})_{\rm phys}$ as $m_{3/2}$ and $T_{RH}$ are varied.  See text for further discussion.  
}
\end{center}
\end{figure}

%===========
The observed value of the baryon asymmetry is $(Y_{B})_{\rm phys} \approx 0.89 \times 10^{-10}$ \cite{Ade:2013rta}.  
Since the effective parameter $\beta$ is independent of $m_{3/2}$ and $T_{RH}$, we can plot the the value of $\beta$ which is required to give $Y_B = (Y_B)_{\rm phys}$, shown in \fref{fig:betaphys}. 
If $\beta$ is larger (smaller) than the value shown at a given $m_{3/2}$ and $T_{RH}$ then B-number is overabundant (underabundant).  
Inspecting the figure reveals that if we hope to satisfy both constraints \eref{eq:m32bound} and \eref{eq:TRHbound}, then we must have $\beta > \beta_{\rm min}$ with 
\begin{align}\label{eq:betabound}
	\beta_{\rm min} \approx 10^{-4} 
\end{align}
in order for gravitino leptogenesis to be successful.  
Alternatively, requiring $\beta < 1$ yields a lower bound on the reheat temperature $T_{RH} > 10^{12} \GeV$, which then provides a firm limit on gravitino leptogenesis scenarios of the type we consider here.

%===========================
% Baryo/Leptogenesis Channels
%===========================
\section{L-number Violating Gravitino Decay Channels}\label{sec:LG_Channels}

%===========
The MSSM admits four operators in the superpotential that violate R-parity.  
One of these four violates baryon number and the remaining three violate lepton number (see \cite{Barbier:2004ez} for a review): 
\begin{align}
	W_{\text{\sc rpv}} &= W_{\text{\sc bv}} + W_{\text{\sc lv}} \label{eq:WRPV} \\
	W_{\text{\sc bv}} &= \frac{1}{2} \lambda^{\prime \prime}_{ijk} \hat{U}^c_i \hat{D}^c_j \hat{D}_k^c \label{eq:WBV} \\
	W_{\text{\sc lv}} &= \frac{1}{2} \lambda_{ijk} \hat{L}_i \hat{L}_j \hat{E}_k^c 
	+ \lambda^{\prime}_{ijk} \hat{L}_i \hat{Q}_j \hat{D}_k^c
	+ \mu_i^{\prime} \hat{H}_u \hat{L}_i \label{eq:WLV} \per
\end{align}
Our notation is summarized in \aref{app:Notation}.  
In this section, we will focus on each of the three L-number violating operators, in order to calculate the parameter $\beta = n_L / n_{3/2}$, defined by \eref{eq:def_beta_from_BF}, and to assess whether the critical value $\beta_{\rm min} = 10^{-4}$ can be reached given constraints on the models.  

%===========
The single B-number violating operator, $W_{\BV}$, has already been shown by CR to be able to give rise to the baryon asymmetry of the universe through gravitino decays \cite{Cline:1990bw} (see also \cite{Dimopoulos:1987rk}).  
In order to draw a contrast between the CR mechanism and gravitino leptogenesis we will briefly review the CR gravitino baryogenesis calculation here.  
The details of the calculation can be found in \aref{app:GravitinoBaryogenesis}.  

The B-number violating operator, given by \eref{eq:WBV}, contains $9$ distinct terms after the sum over flavor indices has been performed.  
One may take a conservative approach and assume that only one of these terms is nonzero.  
In particular, the MSSM superpotential may be extended to include the operator
\begin{align}\label{eq:UDD_WBV}
	W_{\BV} = \frac{1}{2} \lambda^{\prime \prime}_{332} \hat{T}^c \hat{B}^c \hat{S}^c \com
\end{align} 
which violates B-number by one unit.  
Since this operator does not involve any first generation quarks, it is not strongly constrained by bounds on neutron oscillations and heavy nuclei decay.  
The other components of $\lambda^{\prime \prime}_{ijk}$ are generated at one-loop order due to flavor violation in the quark mass matrix, but the smallness of the quark mixing angles renders these contributions negligible \cite{Dimopoulos:1987rk}.  

%===========
In this scenario, the baryon asymmetry is generated directly by gravitino decays after weak sphalerons go out of equilibrium.  
Then, the gravitino need only decay before the onset of BBN at $T_d \approx T_{BBN} \simeq \MeV$, which imposes $m_{3/2} \gtrsim 10 \TeV$ \cite{Weinberg:1982zq}.  
The superpotential $W_{\BV}$ gives rises to the trilinear interactions
\begin{align}\label{eq:UDD_LBV}
	\mathcal{L}_{\BV} =\ & 
	- \frac{1}{2} \lambda_{332}^{\prime \prime} \left( t^c b^c \tilde{s}^c + b^c s^c \tilde{t}^c + s^c t^c \tilde{b}^c \right) 
	- \frac{1}{2} A^{\prime \prime}_{332}\lambda^{\prime \prime}_{332} \tilde{t}^c \, \tilde{b}^c \, \tilde{s}^c 
	+ \hc \com
\end{align}
where the corresponding soft SUSY-breaking term is also included.  
Quadrilinear B-number violating operators also arise, but these scalar interactions only contribute to gravitino decay at the two loop order.  

\begin{figure}[t]
\vspace{0.2in}
\begin{center}
%Subir, you'll have to play around with the values of "width" to get it to look right.  
	\raisebox{-0.5\height}{\includegraphics[width=0.22\textwidth]{figures/2a.jpg}}
	\raisebox{-0.5\height}{\includegraphics[width=0.015\textwidth]{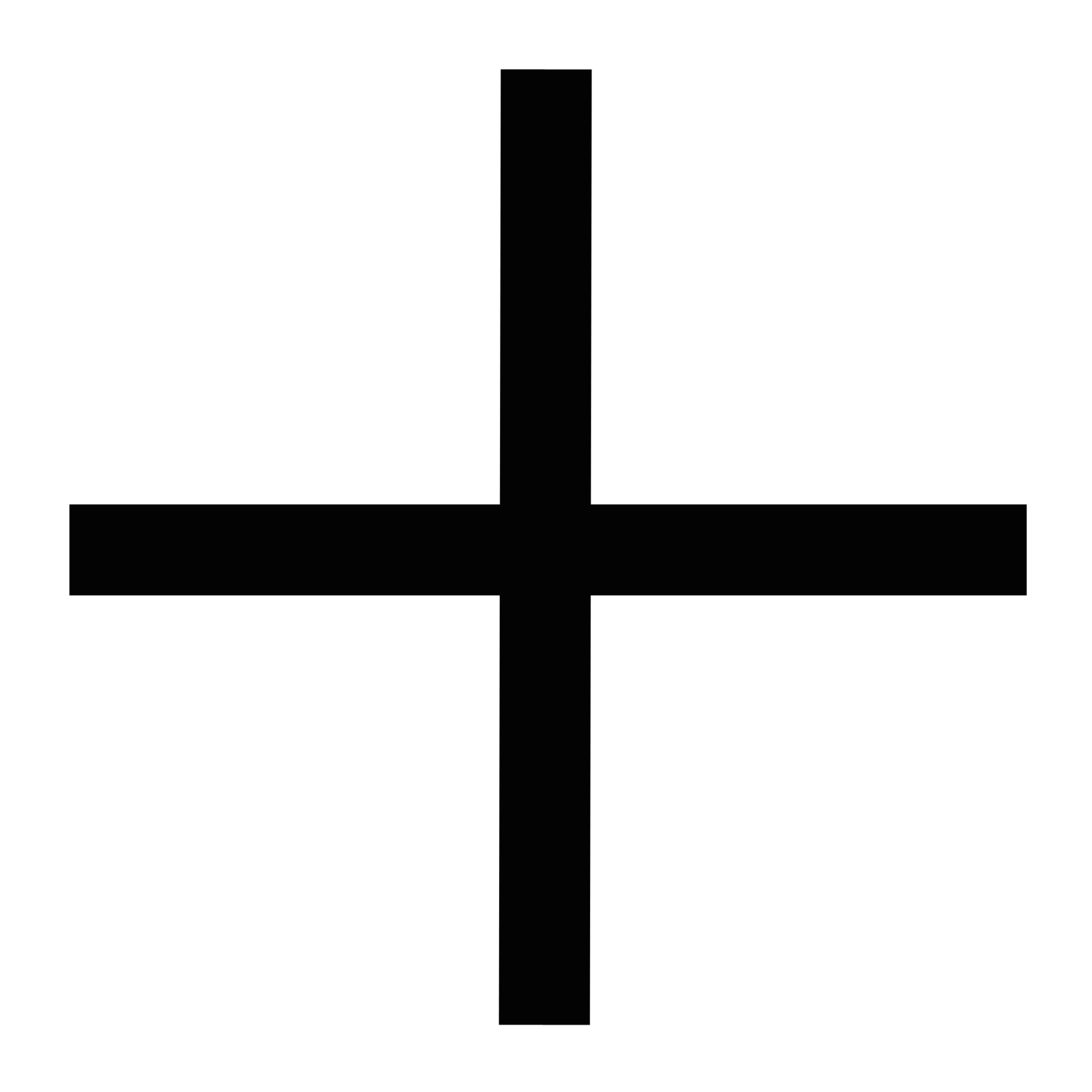}}\ \
	\raisebox{-0.5\height}{\includegraphics[width=0.32\textwidth]{figures/2b.jpg}}\ \
	\raisebox{-0.5\height}{\includegraphics[width=0.015\textwidth]{figures/plus.jpg}}\ \
	\raisebox{-0.5\height}{\includegraphics[width=0.32\textwidth]{figures/2c.jpg}}
\caption{
\label{fig:UDD_CPV}
 Interference between graphs of the form shown here give rise to CP violation via the $\hat{U}^{c}\hat{D}^{c}\hat{D}^{c}$ operator. Additionally graphs with the $\bar{b^c} \tilde{b}^c$ and $\bar{s^c} \tilde{s}^c$ final states are also included.  
}
\end{center}
\end{figure}

%===========
In the CR scenario, baryogenesis can be divided into two stages.  
First, CP-violation biases gravitino decays into anti-squarks ($\bt{q}$) over squarks ($\tilde{q}$), and second, B-number is violated when the anti-squarks decay into quark ($q$) pairs and the squarks decay into anti-quark ($\bar{q}$) pairs.  
In the first stage, the gravitino can decay into the squark directly or by way of a gaugino ($\tilde{X}$).  
In the second stage, the squarks decay through the channels $\tilde{q} \to \bar{q} \bar{q}$ and $\bt{q} \to q q$.  
Thus the two decay chains $\tilde{G} \to q_i \bt{q}_i \to q_i q_j q_k$ and $\tilde{G} \to X \tilde{X} \to X q_i \bt{q}_i \to X q_i q_j q_k$ are responsible for generation of the B-number asymmetry.  
The parameter $\beta = n_{B} / n_{3/2}$ may be calculated from \eref{eq:def_beta_from_BF} with the modification that B-number is counted instead of L-number:  
\begin{align}\label{eq:UDD_beta_def}
	\beta = \sum_{q, \tilde{q}} \Biggl\{ &
	\left( {\rm BR}[ \tilde{G} \to q \bt{q}] - {\rm BR}[ \tilde{G} \to \bar{q} \tilde{q}] \right) \nn
	&+ \sum_{\tilde{X} = \tilde{g}, \tilde{Z}, \tilde{\gamma}} {\rm BR} [ \tilde{G} \to X \tilde{X} ] 
	\left( {\rm BR}[ \tilde{X} \to q \bt{q}] - {\rm BR}[ \tilde{X} \to \bar{q} \tilde{q}] \right)
	\Biggr\} 
	{\rm BR}[ \bt{q} \to qq ]
	\com
\end{align}
where ${\rm BR}$ is the branching ratio of the associated process.  
The gravitino and gaugino decays violate CP due to an interference between graphs of the form shown in \fref{fig:UDD_CPV}.  
The parameter $\beta$ is estimated up to $O(1)$ factors as (see \aref{app:GravitinoBaryogenesis} for details)
\begin{align}\label{eq:UDD_beta_numerical}
	\beta \sim \alpha_{332}^{\prime \prime} \sin \theta_{\CP} \frac{\abs{A_{332}^{\prime \prime}}}{\abs{m_{3/2}}} \frac{42}{N_{\rm eff}} {\rm BR}_{\BV}
	\com
\end{align}
where $\alpha_{332}^{\prime \prime} \equiv \abs{\lambda_{332}^{\prime \prime}}^2 / 4 \pi$ and $N_{\rm eff}$ was given by \eref{eq:Neff_def}, and ${\rm BR}_{\BV} \equiv {\rm BR}[\bt{q} \to qq]$, which may be $O(1)$ if the squark is the LSP.  
The phase $\theta_{\CP} = {\rm Arg}[A_{332}^{\prime \prime} m_{3/2}]$ quantifies the degree of CP violation, which is constrained by the bound on the neutron electric dipole moment (EDM), $d_n \lesssim 10^{-26} \cm$ \cite{Beringer:2012zz}.  
For typical values, 
\begin{align}
	\abs{m_{3/2}} = 20 \TeV
	\quad , \quad
	\abs{A_{332}^{\prime \prime}} = 10 \TeV 
	\quad , \quad
	\alpha_{332}^{\prime \prime} = 0.1 
	\quad , \quad
	\sin \theta_{\CP} = 0.3
\end{align}
one finds \cite{Cline:1990bw}
\begin{align}
	\beta \simeq 0.03
	\qquad {\rm and} \qquad
	d_n \simeq 10^{-26} \cm \per
\end{align}
In this way, a sufficient baryon asymmetry is generated while evading constraints on CP violation from low energy observables.  

%==========

%===========
Successful baryogenesis requires both the generated baryon asymmetry and CP asymmetry to survive potential washout processes.  
The baryon asymmetry could be washed out by the inverse decay processes $qq \to \bt{q}$ and $\bar{q}\bar{q} \to \tilde{q}$, but these processes are suppressed kinematically since $T \approx T_d \ll m_{\tilde{q}}$.  
The CP asymmetry may be washed out by the s-channel scattering with quarks in the plasma, $\tilde{q} \bar{q} \to \tilde{X} \to \bt{q} q$, where a Majorana mass operator is inserted in the gaugino propagator.  
If this process occurs on a time scale shorter than the lifetime of the squarks, then the CP asymmetry would be washed out before they have a chance to decay.  
However, since the CP asymmetry is only carried by the second and third generation squarks, these processes are suppressed by the exponentially low abundances of heavy second and third generation quarks in the $\MeV$-scale plasma.  

We now describe how gravitino leptogenesis, which operates prior to electroweak symmetry breaking, involves qualitatively different constraints in order to remain cosmologically viable and consistent with low energy phenomenology.  

%-------------------------
%  Decay through LLE
%-------------------------
\subsection{Decay through $\hat{L} \hat{L} \hat{E}^c$}\label{sub:tRPV_LV1}

%===========
We now consider the first of the three L-number violating operators which give rise to a lepton asymmetry through gravitino decay through a violation of L-number and R-parity 
\begin{align}\label{eq:LLE_WLV}
	W_{\LV} = \frac{1}{2} \lambda_{233} \hat{L}_2 \cdot \hat{L}_3 \hat{E}_3^c 
\end{align}
The notation ``$\, \cdot \,$'' stands for a contraction of $\SU{2}$ indices with the antisymmetric tensor.  
We have focused on the $233$ component of the tensor $\lambda_{ijk}$ primarily for simplicity.  
As we will discuss below, the high scale of SUSY breaking renders the constraints on $\lambda_{ijk}$ to be very weak.  
The Lagrangian-level interactions are 
\begin{align}\label{eq:LLE_LLV}
	\mathcal{L}_{\LV} = 
	-\frac{1}{2}\lambda_{233} \left(l_{2}\cdot l_{3}\tilde{\tau}^{c}+\tilde{l}_{2} \cdot l_{3}\tau^{c}+l_{2} \cdot\tilde{l}_{3} \tau^{c} \right) 
	-\frac{1}{2} A_{233} \lambda_{233}\tilde{l}_{2}\cdot \tilde{l}_{3}\tilde{\tau}^{c}
	+ \hc \com
\end{align}
which may be compared with \eref{eq:UDD_LBV}.  
Since the electroweak symmetry is still unbroken at the time of gravitino decays, the presence of isospin doublets in \eref{eq:LLE_LLV} simply provides a multiplicative prefactor.  

\begin{figure}[t]
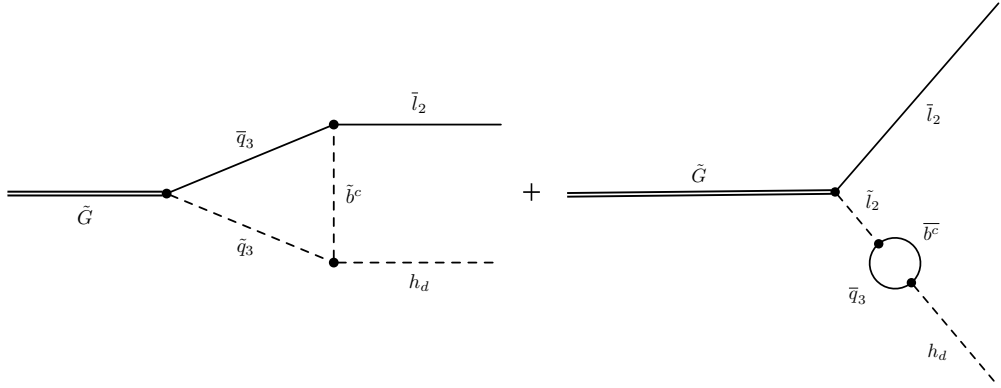

\vspace{0.2in}
\begin{center}
\raisebox{-0.5\height}{\includegraphics[width=0.4\textwidth]{figures/3a.jpg}}\ \ \
	\raisebox{-0.5\height}{\includegraphics[width=0.015\textwidth]{figures/plus.jpg}}\ \ \ \
	\raisebox{-0.5\height}{\includegraphics[width=0.35\textwidth]{figures/3b.jpg}}
\caption{
\label{fig:LLE_HL}
Lepton asymmetry generated directly from a gravitino decay through a loop process. Interference between the two graphs gives rise to CP violation.
}
\end{center}
\end{figure}

%===========
In standard leptogenesis, the Majorana neutrino decays into a lepton and a Higgs in a CP violating manner \cite{Buchmuller:2005eh}.  
Naively one would expect the corresponding supersymmetric decay channels, $\tilde{G} \to \bar{h}_d l_i, \tb{h}_d \tilde{l}_i, h_u l_i$ and $\tilde{h}_u \tilde{l}_i$, to yield the dominant contribution to the lepton asymmetry.  
However, the absence of a direct gravitino--lepton--Higgs vertex requires these decays to be loop suppressed.  
Specifically, graphs of the form shown in \fref{fig:LLE_HL} are responsible for mediating these decays.  
Not only are these decays doubly-loop suppressed, but additionally they require factors of the lepton Yukawa couplings.  
Even the largest Yukawa coupling gives $O(m_{\tau}^2 / v^2) \sim O(10^{-4})$.  
This suppresses these channels compared to the three-body final states that we consider below, and moreover makes them irrelevant for leptogenesis in light of the requirement $\beta_{\rm min} \approx 10^{-4}$ (\eref{eq:betabound}).  

%===========

\begin{figure}[t]
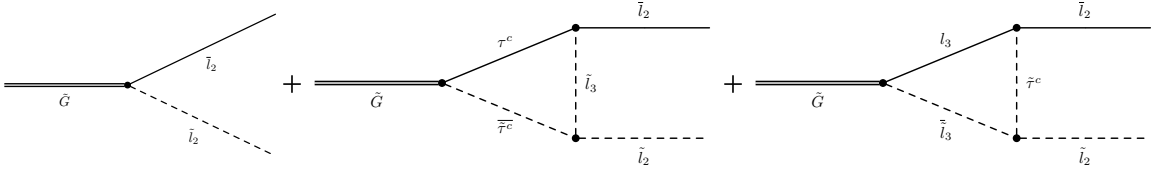

\vspace{0.2in}
\begin{center}
%Subir, you'll have to play around with the values of "width" to get it to look right.  
	\raisebox{-0.5\height}{\includegraphics[width=0.22\textwidth]{figures/4a.jpg}}
	\raisebox{-0.5\height}{\includegraphics[width=0.015\textwidth]{figures/plus.jpg}}\ \
	\raisebox{-0.5\height}{\includegraphics[width=0.32\textwidth]{figures/4b.jpg}}\ \
	\raisebox{-0.5\height}{\includegraphics[width=0.015\textwidth]{figures/plus.jpg}}\ \
	\raisebox{-0.5\height}{\includegraphics[width=0.32\textwidth]{figures/4c.jpg}}
\caption{
\label{fig:LLE_CPV}
Feynman graphs that yield CP violation in gravitino decays via the $\hat{L}\hat{L}\hat{E}^{c}$ operator.  Other graphs with $\bar{l}_3 \tilde{l}_3$ and $\tilde{\tau}^c \overline{\tau^c}$ on the external lines are not shown. 
}
\end{center}
\end{figure}

%===========
The calculation of the appropriate gravitino decay channels then runs parallel to the B-number violating decay that was discussed in \sref{sub:tRPV_LV1}.  
We generate the lepton asymmetry in two stages.  
First, the gravitino decays out of equilibrium through the channels $\tilde{G} \to l \bt{l}$ and $\bar{l} \tilde{l}$ where $l(\bar{l})$ is a (anti-)lepton and $\tilde{l}(\bt{l})$ is an (anti-)slepton.  
This creates equal and opposite CP asymmetries in the leptons and sleptons.  
The heavy sleptons then decay through L-number violating interaction as $\tilde{l} \to ll$ and $\bt{l} \to \bar{l} \bar{l}$, thereby creating the lepton asymmetry.  
We calculate $\beta$ by summing over the final states
\begin{align}
	\beta = \epsilon_{l_2} + \epsilon_{l_3} + \epsilon_{\tau^c}
\end{align}
where
\begin{align}
	\epsilon_{l_2} = {\rm BR}[ \bt{l}_{2} \to {l}_{3} \tau^c] \Biggl\{ &
	\left( {\rm BR}[ \tilde{G} \to {l}_{2} \bt{l}_{2}] - {\rm BR}[ \tilde{G} \to \bar{l}_{2} \tilde{l}_{2}] \right) \nn
	&+ \sum_{\tilde{X} = \tilde{B}, \tilde{W}} {\rm BR} [ \tilde{G} \to X \tilde{X} ] 
	\left( {\rm BR}[ \tilde{X} \to {l}_{2} \bt{l}_{2}] - {\rm BR}[ \tilde{X} \to \bar{l}_{2} {\tilde{l}}_{2}] \right)
	\Biggr\} \com \nn
	\epsilon_{l_3} = {\rm BR}[ \bt{l}_{3} \to {l}_{2} {\tau^c}] \Biggl\{ &
	\left( {\rm BR}[ \tilde{G} \to {l}_{3} \bt{l}_{3}] - {\rm BR}[ \tilde{G} \to \bar{l}_{3} \tilde{l}_{3}] \right) \nn
	&+ \sum_{\tilde{X} = \tilde{B}, \tilde{W}} {\rm BR} [ \tilde{G} \to X \tilde{X} ] 
	\left( {\rm BR}[ \tilde{X} \to {l}_{3} \bt{l}_{3}] - {\rm BR}[ \tilde{X} \to \bar{l}_{3} \tilde{l}_{3}] \right)
	\Biggr\} \com \nn
	\epsilon_{\tau^c} = {\rm BR}[ \bt{\tau}^{c} \to l_{2} l_{3}] \Biggl\{ &
	\left( {\rm BR}[ \tilde{G} \to \tau^c \bt{\tau}^{c}] - {\rm BR}[ \tilde{G} \to \bar{\tau^{c}} \tilde{\tau}^{c}] \right) \nn
	&+ \sum_{\tilde{X} = \tilde{B}} {\rm BR} [ \tilde{G} \to X \tilde{X} ] 
	\left( {\rm BR}[ \tilde{X} \to \tau^c \bt{\tau}^{c}] - {\rm BR}[ \tilde{X} \to \bar{\tau^{c}} \tilde{\tau}^{c}] \right)
	\Biggr\}
\end{align}
CP violation arises in the standard way from the interference of tree level and one loop graphs as shown in \fref{fig:LLE_CPV}.  
Then, following a similar calculational strategy to that employed in CR (see \aref{app:GravitinoBaryogenesis}), we can estimate the branching fractions and obtain 
\begin{align}\label{eq:LLE_beta_numerical}
	\beta \sim \alpha_{233} \sin \theta_{\CP} \frac{1}{N_{\rm eff}} {\rm Max} \left[ 5 \frac{\abs{A_{233}}}{\abs{m_{3/2}}} \, , \, 11\frac{\abs{A_{233}}}{\abs{m_{\tilde{X}}}} \right] {\rm BR}_{\LV}
	\com
\end{align}
%Gravitino = 5 = 2(l2) + 2(l3) + 1(tauc)
%Bino = 5 = 2(l2) + 2(l3) + 1(tauc)
%Wino = 2 = 1(l2) + 1(l3) + 0(tauc)
%Gaugino Total = 11 = 1(bino)*5 + 3(wino)*2
where $\alpha_{233} \equiv \abs{\lambda_{233}}^2 / 4 \pi$, and where  
we have assumed a common mass $m_{\tilde{X}} = m_{\tilde{B}} = m_{\tilde{W}}$ for the binos and winos, and also  for simplicity, we assume a comparable amount of CP violation arises in the gravitino and gaugino decays, i.e., $\theta_{\CP} = {\rm Arg}[A_{233} m_{3/2}] = {\rm Arg}[A_{233} m_{\tilde{X}}]$.  
If it is kinematically forbidden for the sleptons to decay into gauginos or higgsinos, then the branching ratio for the L-number violating decay can be large:
\begin{align}
	{\rm BR}_{\LV} = {\rm BR}[\bt{l}_2 \to l_3 \tau^c] \approx {\rm BR}[\bt{l}_3 \to l_2 \tau^c] \approx {\rm BR}[\bt{\tau}^c \to l_2 l_3] = O(1) \per
\end{align}

%===========
The CP asymmetry in squarks may be washed out by scatterings with leptons in the plasma mediated by a gaugino, i.e., $\tilde{l} \bar{l} \to \tilde{X} \to l \bt{l}$ where $\tilde{X}$ is a bino or a wino.  
In the model of gravitino baryogenesis \cite{Cline:1990bw} reviewed above, such scattering processes were negligible due to the exponential Boltzmann suppression of heavy quarks in the $T \sim \MeV$ scale plasma.  
In the case of leptogenesis, all of the leptons are relativistic at the time of gravitino decay at $T \sim 10^{3} \GeV$, and we must verify that this scattering is out of equilibrium.  
The cross-section for CP violating scattering may be estimated as 
\begin{align}
	\sigma_{\tilde{l}\bar{l}\rightarrow \tilde{\bar{l}}l}\approx\frac{\alpha^2}{m_{\tilde{X}}^2}
\end{align}
where $m_{\tilde{X}}$ is the gaugino mass and $\alpha$ is the fine structure constant.  
Since all leptons are relativistic at this time their equilibrium number density is $n_{l}\sim T^3$ and the CP-violating rate is 
\begin{align}
	\Gamma_{\tilde{l}\bar{l}\rightarrow \tilde{\bar{l}}l} \approx \frac{\alpha^2}{m_{\tilde{X}}^2} T^3 \per
\end{align}
This must be compared against the slepton decay rate
\begin{align}
	\Gamma_{\tilde{l}\rightarrow ll}\approx \alpha_{233}m_{\tilde{l}} 
\end{align}
where $m_{\tilde{l}}$ is the slepton mass.  
The requirement that slepton decay is more rapid that the CP violating scattering imposes a lower bound on the gaugino mass
\begin{align}\label{LLE_Washout}
	m_{\tilde{X}} > \sqrt{ \frac{\alpha^2T^3}{\alpha_{233}m_{\tilde{l}}} } \per
\end{align}
Taking $T \sim 10^{3} \GeV$ to be the temperature at which gravitinos decay, $\alpha \sim 10^{-2}$, $\alpha_{233} = O(1)$, and $m_{\tilde{l}} \sim m_{3/2} \sim 10^{8} \GeV$ we obtain the weak bound $m_{\tilde{X}} \gtrsim O(1 \GeV)$.  
For the scale that we are considering however, we expect $m_{\tilde{X}} \sim 10^{8} \GeV$, so wash out is not a problem.  

%===========
Low energy observables can be used to constrain CP violation in baryogenesis models.  
In the decoupling limit, in which the SUSY breaking scale is taken to infinity, these constraints disappear.  
Although the SUSY breaking scale that we consider is high, we still need to check that it is sufficiently high so that phenomenologically unacceptable corrections to low energy observables are negligible.  

CP violation in the lepton sector generates an electron EDM, $d_{e}$.  
The contribution to $d_e$ in the MSSM may be estimated as \cite{Abel:1999yz}
\begin{align}
	d_{e} \approx \frac{m_{e}}{m_{\tilde{l}}^2}\sin\theta_{\CP}
\end{align}
where $m_{\tilde{l}}$ is the slepton mass, $m_{e}$ is the electron mass, and $\theta_{\CP}$ is the CP violating phase.  
The current bound, $d_{e} < 5\times10^{-14} \GeV^{-1}$ \cite{Beringer:2012zz}, imposes a constraint on the slepton mass $m_{\tilde{l}} \gtrsim 10^{5} \GeV$, which is easily accommodated for our fiducial superpartner mass scale $m_{3/2} \sim 10^{8} \GeV$.  
The interactions in \eref{eq:LLE_LLV} also violate lepton flavor, which is constrained by bounds on the decay $\mu\rightarrow e\gamma$.  
The branching ratio may be estimated as 
\begin{align}
	{\rm BR}[\mu^{-} \to e^{-} \gamma] \sim \alpha_{233} \frac{m_{\mu}^2}{m_{\tilde{l}}^2} \per
\end{align}
The current bound, ${\rm BR}[\mu^{-} \to e^{-} \gamma] \lesssim 2.4\times10^{-12}$ \cite{Beringer:2012zz}, imposes a lower bound on the slepton mass, $m_{\tilde{l}} \gtrsim O(10^{4} \GeV)$, which is also easy to accommodate.

%==========

%-------------------------
%  Decay through LQD
%-------------------------
\subsection{Decay through $\hat{L} \hat{Q} \hat{D}^c$}\label{sub:tRPV_LV2}

%==========
The symmetries and field content of the MSSM admit just one other trilinear R-parity and L-number violating operator. Once again we will focus on a single element of the flavor tensor and write the R-parity violating superpotential as 
\begin{align}
	W_{\LV} = \frac{1}{2}\lambda^{\prime}_{233} \hat{L}_2 \cdot \hat{Q}_3 \hat{D}_3^c \com
\end{align}
which violates L-number but preserves B-number. The interaction Lagrangian contains the following terms
\begin{align}\label{eq:LQD_LLV}
	\mathcal{L}_{\LV} = -\frac{1}{2}\lambda_{233}^{\prime} \left( l_{2}\cdot q_{3}\tilde{b}^{c}+\tilde{l}_{2}\cdot q_{3}b^{c}+l_{2}\cdot\tilde{q}_{3}b^{c}\right)-\frac{1}{2}A_{233}\lambda_{233}^{\prime}\tilde{l}_{2}\cdot\tilde{q}_{3}\tilde{b}^{c}+\hc \per
\end{align}
which may be compared with \eref{eq:LLE_LLV}.  

%==========
The calculation of $\beta$ parallels the discussion in \sref{sub:tRPV_LV1}.  
The qualitative difference is that in the first stage of leptogenesis, the gravitino can decay into either a lepton--slepton pair or a quark--squark pair.  
Subsequently, both the slepton and the squark decay violating L-number.  
Summing over the various decay channels, we obtain 
\begin{align}
	\beta \equiv \epsilon_{l_2} + \epsilon_{q_3} + \epsilon_{b^c}
\end{align}
where 
\begin{align}
	\epsilon_{l_2} = {\rm BR}[ \bt{l}_{2} \to q_{3} b^{c}] \Biggl\{ &
	\left( {\rm BR}[ \tilde{G} \to l_{2} \bt{l}_{2}] - {\rm BR}[ \tilde{G} \to \bar{l}_{2} \tilde{l}_{2}] \right) \nn
	&+ \sum_{\tilde{X} = \tilde{B}, \tilde{W}} {\rm BR} [ \tilde{G} \to X \tilde{X} ] 
	\left( {\rm BR}[ \tilde{X} \to {l}_{2} \bt{l}_{2}] - {\rm BR}[ \tilde{X} \to \bar{l}_{2} \tilde{l}_{2}] \right)
	\Biggr\} \com \nn
	\epsilon_{q_3} = {\rm BR}[ \bt{q}_{3} \to {l}_{2} {b}^{c}] \Biggl\{ &
	\left( {\rm BR}[ \tilde{G} \to {q}_{3} \bt{q}_{3}] - {\rm BR}[ \tilde{G} \to \bar{q}_{3} \tilde{q}_{3}] \right) \nn
	&+ \sum_{\tilde{X} = \tilde{B}, \tilde{W}, \tilde{g}} {\rm BR} [ \tilde{G} \to X \tilde{X} ] 
	\left( {\rm BR}[ \tilde{X} \to {q}_{3} \bt{q}_{3}] - {\rm BR}[ \tilde{X} \to \bar{q}_{3} \tilde{q}_{3}] \right)
	\Biggr\} \com \nn
	\epsilon_{b^c} = {\rm BR}[ \bt{b}^{c} \to {l}_{2} {q}_{3}] \Biggl\{ &
	\left( {\rm BR}[ \tilde{G} \to {b}^{c} \bt{b^c}] - {\rm BR}[ \tilde{G} \to \bar{b^c} \tilde{b}^{c}] \right) \nn
	&+ \sum_{\tilde{X} = \tilde{B}, \tilde{g}} {\rm BR} [ \tilde{G} \to X \tilde{X} ] 
	\left( {\rm BR}[ \tilde{X} \to {b}^{c} \bt{b^c}] - {\rm BR}[ \tilde{X} \to  \bar{b^c} \tilde{b}^{c}] \right)
	\Biggr\}  \per
\end{align}
Once again using the results of \aref{app:GravitinoBaryogenesis} we estimate 
\begin{align}\label{eq:LQD_beta_numerical}
	\beta \sim \alpha_{233}^{\prime} \sin \theta_{\CP}  \frac{1}{N_{\rm eff}} {\rm Max} \left[ 11\frac{\abs{A_{233}}}{\abs{m_{3/2}}} \, , \, 27\frac{\abs{A_{233}}}{\abs{m_{\tilde{X}}}} \right] {\rm BR}_{\LV}
	\com
\end{align}
%Gravitino = 11 = 2(l2) + 6(q3) + 3(bc)
%Bino = 5 = 2(l2) + 6(q3) + 3(bc)
%Wino = 2 = 1(l2) + 3(q3) + 0(bc)
%Gluino = 2 = 0(l2) + 1(q3) + 1(bc)
%Gaugino Total = 27 = 1(bino)*5 + 3(wino)*2 + 8(gluino)*2
where $\alpha_{233}^{\prime} \equiv \abs{\lambda_{233}^{\prime}}^2 / 4 \pi$.  
With the appropriate spectral constraints we can obtain 
\begin{align}
	{\rm BR}_{\LV} = {\rm BR}[ \bt{l}_{2} \to q_{3} b^{c}] \approx 
	{\rm BR}[ \bt{q}_{3} \to {l}_{2} {b}^{c}] \approx 
	{\rm BR}[ \bt{b}^{c} \to {l}_{2} {q}_{3}] \approx O(1) \com
\end{align}
as in the previous cases.  

\begin{figure}[t]
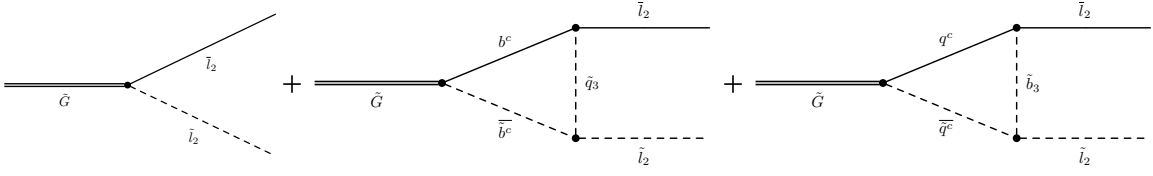

\vspace{0.2in}
\begin{center}
%Subir, you'll have to play around with the values of "width" to get it to look right.  
	\raisebox{-0.5\height}{\includegraphics[width=0.22\textwidth]{figures/5a.jpg}}
	\raisebox{-0.5\height}{\includegraphics[width=0.015\textwidth]{figures/plus.jpg}}\ \
	\raisebox{-0.5\height}{\includegraphics[width=0.32\textwidth]{figures/5b.jpg}}\ \
	\raisebox{-0.5\height}{\includegraphics[width=0.015\textwidth]{figures/plus.jpg}}\ \
	\raisebox{-0.5\height}{\includegraphics[width=0.32\textwidth]{figures/5c.jpg}}
\caption{
\label{fig:LQD_CPV}
Feynman graphs that yield CP violation in gravitino decays via the $\hat{L}\hat{Q}\hat{D}^{c}$ operator.  Other graphs with $\bar{q}_3 \tilde{q}_3$ and $\tilde{b}^c \overline{b^c}$ in the final state are not shown.  
}
\end{center}
\end{figure}

%==========
Apart from the distinctions discussed thus far, the remainder of the analysis of this case follows similarly to \sref{sub:tRPV_LV1}.  
Washout is possible due to s-channel scatterings through gauginos, but the avoidance of washout imposes only a very weak bound on the gaugino mass.  
Empirical constraints, arising from electron and neutron EDMs and lepton flavor violation, have little constraining power on the R-parity violating couplings, $\lambda'_{233}$ and $A_{223}^{\prime}$, due to the high scale of SUSY breaking.  

%==========

%-------------------------
%  Decay through HL
%-------------------------
\subsection{Decay through $\hat{H}_u \hat{L}$}\label{sub:bRPV_LV}

%==========
As a last case we will consider the bilinear R-parity and L-number violating operator, 
\begin{align}\label{eq:HL_WLV}
	W_{\LV} = \mu_i^{\prime} \hat{H}_u \cdot \hat{L}_i \per
\end{align}
This operator supplements the R-parity symmetric terms from the MSSM
\begin{align}
	W_{\MSSM} = \mu \hat{H}_u \cdot \hat{H}_d + (\lambda^{e})_{ij} \hat{H}_d \cdot \hat{L}_i \hat{E}^c_{j} - (\lambda^{u})_{ij} \hat{H}_u \cdot \hat{Q}_i \hat{U}^c_{j} + (\lambda^{d})_{ij} \hat{H}_d \cdot \hat{Q}_i \hat{D}^c_{j} \per
\end{align}
The full superpotential $W = W_{\MSSM} + W_{\LV}$ yields the Lagrangian $\mathcal{L} = \mathcal{L}_{\LV}^{\rm bi} + \mathcal{L}_{\LV}^{\rm tri} + \mathcal{L}_{\LV}^{\rm quad} + \mathcal{L}_{\LP} + \mathcal{L}_{\MSSM}$ where 
\begin{align}\label{eq:HL_LLV}
	-\mathcal{L}_{\LV}^{\rm bi} &= \mu_{i}^{\prime} \tilde{h}_{u}\cdot l_{i} + B^{u}_{i}h_{u}\cdot\tilde{l}_{i} + \left( B^{d}_{i} + \mu^{\ast} \mu_{i}^{\prime} \right) h_{d}^{\dagger}\tilde{l}_{i} +\hc \com \\
	-\mathcal{L}_{\LV}^{\rm tri} &= -\mu_{i}^{\prime \, \ast} (\lambda^{u})_{jk} \, \tilde{l}_{i}^{\dagger} \tilde{q}_{j} \tilde{u^c}_{k} + \mu_{i}^{\prime \ast} (\lambda^{e})_{ij} \, h_{u}^{\dagger} h_{d} \tilde{e^c}_{j} + \hc \com \qquad {\rm and} \\
	-\mathcal{L}_{\LP} &= \abs{\mu_i^{\prime}}^2 h_u^{\dagger} h_u + (\mu_i^{\prime \, \ast} \mu_j^{\prime}) \tilde{l}_i^{\, \dagger} \tilde{l}_j
\end{align}
are the bilinear L-violating, trilinear L-violating, and L-preserving contributions that are in addition to the MSSM Lagrangian, $\mathcal{L}_{\MSSM}$. 
We will not need the quadrilinear terms, $\mathcal{L}_{\LV}^{\rm quad}$, since they only contribute to gravitino decay at the two loop order.   
As we have done in the previous sections, we will suppose that $W_{\LV}$ is the only source of R-parity violation at tree-level.  

%==========
We will see that bilinear L-number violation provides various gravitino decay channels that generate a lepton asymmetry.  
This scenario, however, is significantly constrained, because the mixings in \eref{eq:HL_LLV} allow L-number violation to enter into low energy observables, specifically the neutrino mass, at tree level.  
This is in contrast to the previous cases of trilinear R-parity violation in which L-number violating effects were loop suppressed.  
If the mass scale of the sleptons and neutralinos is comparable to the fiducial gravitino mass that we have considered in the previous sections, $m_{3/2} \sim 10^{8} \GeV$, then the neutrino mass constraints bound the mixings so as to forbid the generation of a sufficiently large lepton asymmetry.  
Only if the mass scale is lifted to the somewhat more uncomfortable scale, $m_{3/2} \gtrsim O(10^{10} - 10^{11} \GeV)$, can the low energy constraints be evaded.   While it makes this leptogenesis scenario less attractive, for completeness we review the parameter ranges that remain viable for this case as well. 

%==========
In this scenario, the lepton asymmetry is created by the decays of the gravitino or gaugino into a lepton and a Higgs boson.  
Depending on the spectrum, the gravitino could also decay into a slepton and a Higgsino through the R-parity violating mixing.  
However, the slepton and Higgsino would eventually decay back through the R-parity violating operator into SM particles, and this may lead to significant washout of the lepton asymmetry.  
Therefore, we assume the spectrum
\begin{align}\label{eq:HL_spectrum}
	m_{\tilde{l}} \sim m_{\tilde{h}} \sim m_{\tilde{q}} \gtrsim m_{3/2} \gtrsim m_{\tilde{X}} \sim m_{h} \gg m_{l} \, , \, m_{q} \com
\end{align}
and we focus on the two decay channels (and their CP conjugates), $\tilde{V} \to l_{i} \bar{h}_d$ and $l_{i} h_u$, where $\tilde{V} = \tilde{G}, \tilde{B},$ or $\tilde{W}$.  
In the case of trilinear L-number violation, we had dismissed these two-body final states as subdominant (see \sref{sub:tRPV_LV1}) because they arose from an interference of two one-loop graphs, but for bilinear L-number violation the decays can proceed due to the tree level mixing.  

\begin{figure}[t]
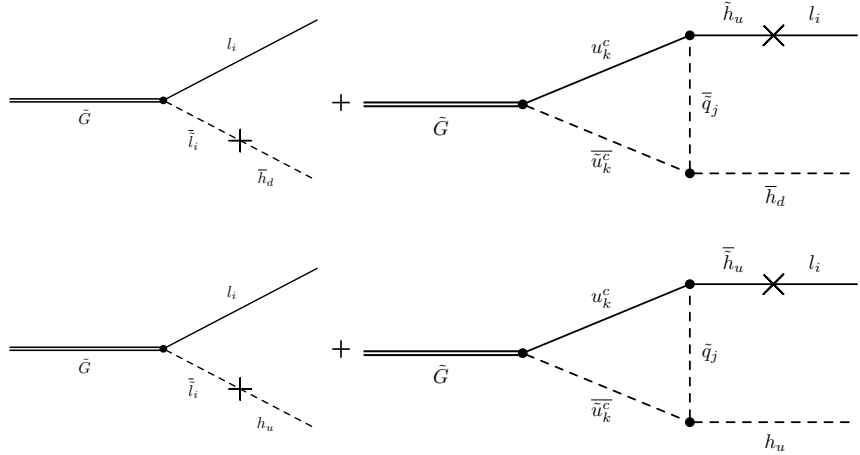

\vspace{0.2in}
\begin{center}
	\raisebox{-0.5\height}{\includegraphics[width=0.25\textwidth]{figures/6a.jpg}}\ \
	\raisebox{-0.5\height}{\includegraphics[width=0.015\textwidth]{figures/plus.jpg}}\ \
	\raisebox{-0.5\height}{\includegraphics[width=0.4\textwidth]{figures/6b.jpg}}\\
	\vspace{5mm}
	\raisebox{-0.5\height}{\includegraphics[width=0.25\textwidth]{figures/6c.jpg}}\ \
	\raisebox{-0.5\height}{\includegraphics[width=0.015\textwidth]{figures/plus.jpg}}\ \
	\raisebox{-0.5\height}{\includegraphics[width=0.4\textwidth]{figures/6d.jpg}}
\caption{
\label{fig:HL_CPV}
Examples of the Feynman graphs whose interference give rise to a lepton asymmetry via bilinear R-parity and L-number violation.
}
\end{center}
\end{figure}

%==========
For this scenario, the lepton asymmetry is given by summing over the two final states
\begin{align}
%	\beta = ( \eps_{l_i \bar{h}_d} + \eps_{\tilde{l}_i \tb{h}_d} + \eps_{l_i h_u} + \eps_{\tilde{l}_i \tilde{h}_u} ) f_{wo} 
	\beta = ( \eps_{l_i \bar{h}_d} + \eps_{l_i h_u} ) f_{wo} 
\end{align}
where
\begin{subequations}\label{eq:HL_epsilon_def}
\begin{align}
	\eps_{l_i \bar{h}_d} 
	&\equiv \sum_{i} \Bigl\{ \big({\rm BR} [ \tilde{G} \to l_{i} \bar{h}_{d}  ] - {\rm BR} [ \tilde{G} \to \bar{l}_{i} {h}_{d}  ] \big) \nn
	& \qquad \quad + \sum_{\tilde{X} = \tilde{B}, \tilde{W} } {\rm BR}[\tilde{G} \to X \tilde{X}] \big({\rm BR} [ \tilde{X} \to l_{i} \bar{h}_{d}  ] - {\rm BR} [ \tilde{X} \to \bar{l}_{i} {h}_{d}  ] \big) \Bigr\} \\
	\eps_{l_i h_u} 
	&\equiv \sum_{i} \Bigl\{ \big({\rm BR} [ \tilde{G} \to l_{i} h_{u}  ] - {\rm BR} [ \tilde{G} \to \bar{l}_{i} \bar{h}_{u}  ] \big) \nn
	& \qquad \quad + \sum_{\tilde{X} = \tilde{B}, \tilde{W} } {\rm BR}[\tilde{G} \to X \tilde{X}] \big({\rm BR} [ \tilde{X} \to l_{i} h_{u}  ] - {\rm BR} [ \tilde{X} \to \bar{l}_{i} \bar{h}_{u} ] \big) \Bigr\} 
\end{align}
\end{subequations}
and $f_{wo} \leq 1$ is a suppression factor to account for washout effects (see below).  
CP violation arises from the interference of graphs such as the ones shown in \fref{fig:HL_CPV}.  
We could include additional graphs with more insertions of the mixing operators, but as we will see below the mixing induced by the parameters $\mu_{i}^{\prime}$, $B_i^{d}$, and $B_{i}^{u}$ must be small compared to the superpartner mass scale, and the higher order graphs can be neglected from a perturbative standpoint.  
Additionally, graphs containing factors of the electron and down-type quark Yukawa coupling are subdominant.    
Taking $\abs{B^u} \sim \abs{B^d} \sim \abs{\mu^{\prime}}^2$ and the spectrum in \eref{eq:HL_spectrum} for simplicity, we obtain the order of magnitude estimate 
\begin{align}\label{eq:HL_beta_estimate}
	\beta \sim \abs{\lambda^u}^2 \sin \theta_{\CP} \frac{1}{N_{\rm eff}} \frac{\abs{B^d} \abs{\mu^{\prime}} }{\abs{m_{3/2}}^3} f_{wo}
\end{align}
where $\theta_{\CP} = {\rm Arg}[ B^d \mu^{\prime \ast} m_{3/2}]$.  

%==========
The magnitude of the washout depends critically on the spectrum of the superpartners.  
For instance, if the gravitino can decay into squarks, then these will scatter on quarks in the plasma and potentially violate R-parity and L-number.  
It is also possible for two SM particles to scatter violating R-parity, but since the energy of the plasma ($\sim \TeV$) is insufficient to produce a heavy superpartner on shell, such a scattering must contain two factors of the mixing, which is highly suppressed.  
Thus, if we assume that the gravitino cannot decay into on shell squarks and sleptons, then washout can be negligible.  

%==========

%==========
Let us now turn to the low energy constraints on this model.  
The neutrinos mix with the up-type Higgsino through the bilinear operator $\mathcal{L}_{\LV}^{\rm bi} \ni \mu_{i}^{\prime} \tilde{h}_u^0  \nu_i$.  
This mixing causes the neutrinos to acquire a mass, and therefore neutrino mass constraints impose an upper bound on the mixing \cite{Hall:1983id, Brahm:1990xx}.  
The neutralinos can be integrated out to yield a neutrino mass \cite{Joshipura:1994ib}
\begin{align}\label{eq:HL_mnu}
	m_{\nu} \sim \frac{m_{Z}^2\abs{\mu^{\prime}}^2}{m_{\tilde{\chi}}^3} \frac{1}{1 + t_{\beta}^2} \per
\end{align}
where $m_{\tilde{\chi}} \gg m_Z , m_{\nu}$ is the neutralino mass and $t_{\beta} = \langle h_u \rangle / \langle h_d \rangle \sim O(1)$.  
Taking $m_{\nu} \lesssim 1 \eV$ and the fiducial reference $m_{\tilde{\chi}} \sim m_{3/2}$, we obtain the bound 
\begin{align}\label{eq:mu_bound}
	\frac{\abs{\mu^{\prime}}}{m_{3/2}} \lesssim 10^{-3} \sqrt{ \frac{m_{3/2}}{10^{8} \GeV}} \ \sqrt{ \frac{1}{1 + t_{\beta}^2} } \per
\end{align}
Despite the high scale of SUSY breaking, a strong constraint is obtained because the neutrino mass arises at tree level, unlike in the cases of the trilinear operators.  

After electroweak symmetry breaking the Higgs fields acquire vacuum expectation values and the mixings in \eref{eq:HL_LLV} induce tadpole terms for the sneutrino fields
\begin{align}
	t_{\tilde{\nu}_i} =  (B^{d}_i + \mu^{\ast} \mu_i^{\prime}) \, v_{d} - B^{u}_i \, v_u \per
\end{align}
This causes the sneutrinos to acquires VEVs, which may be estimated as 
\begin{align}\label{eq:sneutrino_VEV}
	v_{\tilde{\nu}} \approx \frac{(B^{d}+\mu^{*}\mu^{\prime})\cos\beta-B^{u}\sin\beta}{M_{\tilde{l}}^2+\abs{\mu^{\prime}}^2 } \, v 
\end{align}
where $M_{\tilde{l}}^2$ is the soft SUSY-breaking slepton mass parameter.  
The charged slepton VEVs are protected by the residual electromagnetic symmetry.  
The VEV $v_{\tilde{\nu}}$ causes the gauginos and neutrinos to mix and gives rise to a neutrino mass \cite{Ellis:1984gi}
\begin{align}
	m_{\nu} \sim \frac{m_Z^2}{v^2} \frac{v_{\tilde{\nu}}^2}{m_{\tilde{X}}}
\end{align}  
where $m_{\tilde{X}}$ is the gaugino mass.  
Once again taking $m_{\tilde{X}} \sim m_{3/2}$, the observed neutrino mass scale implies the bound
\begin{align}\label{eq:snu_bound}
	\frac{v_{\tilde{\nu}}}{m_{3/2}} \lesssim 10^{-8} \sqrt{ \frac{10^{8} \GeV}{m_{3/2}} } \per
\end{align}
Provided that there is not an unnatural tuning in \eref{eq:sneutrino_VEV} which would give $v_{\tilde{\nu}} \ll O(B^d) \sim O(B^u)$, the bound in \eref{eq:snu_bound} imposes
\begin{align}\label{eq:B_bound}
	\frac{B^{d}}{m_{3/2}^2} \sim \frac{B^{u}}{m_{3/2}^2} \lesssim 10^{-3} \sqrt{ \frac{m_{3/2}}{10^{8} \GeV}} \per
\end{align}
The lepton asymmetry, estimated by \eref{eq:HL_beta_estimate}, requires two factors of the mixing, and in light of the constraints Eqs.~(\ref{eq:mu_bound}) and (\ref{eq:B_bound}), it is not possible to achieve $\beta \gtrsim 10^{-4}$, which is required for successful gravitino leptogenesis (see \eref{eq:betabound}).  
The gravitino mass scale must be lifted to at least $m_{3/2} \gtrsim O(10^{10} - 10^{11} \GeV)$ in order for the bounds to be evaded, unless we also impose artificial tuning to lift the masses of the sleptons and neutralinos.  

%==========
Finally, one may wonder why we have not considered L-number violation entering directly in the the K{\"a}hler potential instead of the superpotential.  
The term $K_{\LV} = \alpha_{i} \hat{H}_d^{\dagger} \hat{L}_i$ may be added to the K{\"a}hler potential without violating any gauge symmetries or supersymmetry.  
This term leads to the desired tree level gravitino -- lepton -- Higgs vertex.  
However, since the addition of this term makes the K{\"a}hler potential non-diagonal, it will also result in non-canonical kinetic terms.  
A basis may be found in which the kinetic terms are diagonal by rotating away the bilinear coupling, but this also removes the tree level gravitino--lepton--Higgs vertex.  

%===========================
% Conclusions
%===========================
\section{Conclusions}\label{sec:Conclusions}

%===========
We have considered here an additional possible mechanism for the creation of the baryon asymmetry of the universe via gravitino decays in the MSSM.  
In this scenario, the out of equilibrium decay of the gravitino gives rise to a lepton asymmetry that is subsequently converted into a baryon asymmetry by weak sphalerons.  
The requirements of CP and L-number violation are then provided by three of the MSSM's possible R-parity violating operators:  $W = \hat{L} \hat{L} \hat{E}^c, \hat{L} \hat{Q} \hat{D}^c$, and $\hat{H}_u \hat{L}$.  
For the case of the two trilinear operators, the gravitino decay channels responsible for L-number creation are similar to the B-number creation gravitino decays discussed by \rref{Cline:1990bw} for the operator $W = \hat{U}^c \hat{D}^c \hat{D}^c$, and the analysis of the subsequent generation of lepton number asymmetry follows a similar line of analysis, with some key changes due to the differing presumed scale of gravitino mass.  

For comparison purposes and to demonstrate the viability of these scenarios, we provide in,   
\tref{tab:LLE_LQD} some sample parameter sets which may produce the correct order of magnitude of the observed baryon asymmetry, without coming into conflict with low energy observables such as EDMs and $\mu \to e \gamma$.  

In the case of the bilinear operator $W = \hat{H}_u \hat{L}$, a lepton asymmetry can be generated through mixing between leptons and Higgsinos and between sleptons and Higgs bosons.  
In the particular case where the gravitino decays into a Higgs boson and a lepton via the R-parity violating mixing, the neutrino also acquires a mass by virtue of this mixing.  As a result, bounds on the neutrino mass constrain the mixing to the point that an insufficient baryon asymmetry is generated unless the mass scale of the gravitino  is increased to $m_{3/2} \gtrsim 10^{10-12} \GeV$, as demonstrated in \tref{tab:HL}.  

%===========
All of these mechanisms of gravitino leptogenesis require an unconventional spectrum of superpartners to the SM fields, or equivalently, a restriction on the mechanism of SUSY breaking.  
The gravitino must be heavy, $m_{3/2} \gtrsim 10^{8} \GeV$, to ensure that it decays while the weak sphalerons are still in equilibrium, and this corresponds to a SUSY breaking scale $M_S \gtrsim 10^{13} \GeV$.  
There is no {\it a priori} reason to think that the scale of SUSY-breaking cannot be so high, and in fact, the cosmological consequence of high-scale SUSY breaking have been studied \cite{Pagels:1981ke, Weinberg:1982zq}.  
However, if $M_S \gg \TeV$ then supersymmetry does not provide a natural solution to the Higgs hierarchy problem.  
Nevertheless, the lack of evidence for supersymmetry at the LHC already implies that the scale of SUSY breaking is higher than naturalness arguments would suggest, and it is therefore worth exploring the possibility that it could be {\it much} higher as, for example, in the case of split supersymmetry \cite{ArkaniHamed:2004fb, Giudice:2004tc, Arvanitaki:2012ps}.  
Also, if the gravitino is to be responsible for leptogenesis, it must decay and therefore cannot be the LSP.  Since the latter possibility is a generic prediction of gauge mediated models of SUSY breaking \cite{Giudice:1998bp}, therefore gauge mediation seems incompatible with gravitino leptogenesis.  

%===========
\begin{table}[t]
\hspace{-0.8cm}
%\begin{center}
\begin{tabular}{|c|c|c|c|c|c|c|c||c|c|c|c|}
\hline
$\abs{m_{3/2}}$ & $M_{\tilde{X}}$ & $M_{\tilde{l}} \sim M_{\tilde{q}}$ & $\abs{A}$ & $\theta_{\CP}$ & $\alpha$ & ${\rm BR}_{\LV}$ & ${T}_{RH}$ & $\beta (10^{-3})$ & $\frac{Y_{B}^{\ast}}{Y_{B}^{\rm (obs)}}$ & $\frac{d_{e}}{d_{e}^{\rm (lim)}}$ & $\frac{ {\rm BR}_{\mu \to e \gamma} }{ {\rm BR}^{\rm (lim)}_{\mu \to e \gamma} }$ \\ 
\hline \hline
1 & $0.1$ & 0.1 & 0.01 & 0.4 & 0.1 & 0.9 & $10^{15}$ & $2$ & 1.0 & $4\cdot 10^{-5}$ & $4\cdot 10^{-6}$ \\
1 & $0.01$ & $0.001$ & $0.01$ & 0.01 & 0.1 & 0.9 & $10^{15}$ & $0.6$ & $0.3$ & $10^{-2}$ & $4\cdot 10^{-2}$ \\
1 & $0.01$ & $0.001$ & $0.03$ & 0.1 & 0.01 & 0.9 & $10^{15}$ & $2$ & $0.7$ & $10^{-1}$ & $4\cdot 10^{-3}$ \\
1 & $0.01$ & $0.01$ & $0.03$ & 0.1 & 0.01 & 0.6 & $10^{15}$ & $1$ & $0.5$ & $10^{-3}$ & $4\cdot 10^{-5}$ \\
5 & $1$ & $0.1$ & $0.05$ & $1$ & 0.4 & 0.8 & $10^{14}$ & $10$ & $0.4$ & $8\cdot 10^{-5}$ & $2\cdot10^{-5}$ \\
5 & $0.05$ & $0.1$ & $0.05$ & $1$ & 0.4 & 0.8 & $10^{13}$ & $185$ & $0.7$ & $8\cdot 10^{-5}$ & $2\cdot10^{-5}$ \\
10 & $1$ & $0.01$ & $3.0$ & 0.1 & 0.1 & 0.1 & $10^{15}$ & $2$ & $0.8$ & $10^{-3}$ & $4\cdot 10^{-4}$ \\
$10^4$ & $0.1$ & $10$ & $1$ & $0.05$ & 0.1 & 0.5 & $10^{14}$ & $17$ & $0.7$ & $5\cdot10^{-10}$ & $4\cdot10^{-10}$ \\
\hline
\end{tabular}
%\end{center}
\caption{\label{tab:LLE_LQD}Typical parameter sets for the model $W_{\LV} = \hat{L} \hat{L} \hat{E}^c$. The input parameters are the gravitino mass ($\abs{m_{3/2}}$), the gaugino mass ($M_{\tilde{X}}$), the squark or slepton mass ($M_{\tilde{l}} \sim M_{\tilde{q}}$), the A parameter ($\abs{A}$), the CP-violating phase ($\theta_{\CP} = {\rm Arg}[A m_{3/2}]$), the R-parity violating Yukawa coupling ($\alpha = \abs{\lambda}^2 / 4 \pi$), the L-number violating branching ratio (${\rm BR}_{\LV}$), and the reheat temperature ($T_{RH}$). Dimensionful parameters are expressed in units of $10^{8} \GeV$ except $T_{RH}$ which is in $\GeV$, $\theta_{\CP}$ is in radians. We estimate the baron asymmetry, $Y_B^{\ast}$ using \eref{eq:YB_estimate} along with the approximate estimate for $\beta$ given by \eref{eq:LLE_beta_numerical}, as well as estimate the ratio of the electron EDM ($d_e$) to the observed upper limit upper limit, and the branching ratio for $\mu \to e \gamma$ with respect to its observed upper limit. The results for $\hat{L} \hat{Q} \hat{D}^c$ would be very similar, differing only by an O(1) factor.
}
\end{table}%

%===========
While our consideration of gravitino leptogenesis is motivated in part by neutrino moderated leptogenesis, there  are a number of key differences.  
The gravitino always decays out of equilibrium by virtue of its universal gravitational strength coupling, whereas the Majorana neutrino decay will be accompanied by some washout factor due to inverse decays \cite{Buchmuller:2005eh}.  
In gravitino decay, the violation of L-number and CP are a consequence of the MSSM's R-parity violating operators.  
On the other hand, the Majorana neutrino mass operator violates L-number and the mass matrix carries the CP-violating phases.  
Successful gravitino leptogenesis requires a high SUSY-breaking scale, $M_S \gtrsim 10^{13} \GeV$, while Majorana neutrino leptogenesis requires a high Majorana mass scale, $M_{N_R} \gtrsim 10^{10} \GeV$. Since these scales are much higher than the energies accessible in the laboratory today, conventional low energy tests of CP-violation do not probe the high energy CP-violating parameters that may be responsible for generation of the lepton asymmetry (with few exceptions in cases of standard neutrino leptogenesis \cite{Frampton:2002qc} \cite{Endoh:2002wm}).
Similarly, bounds on lepton flavor violation in the form of the process $\mu \to e \gamma$ are insensitive to the L-number violation that is responsible for leptogenesis.

\begin{table}[t]
\begin{center}
\begin{tabular}{|c|c|c|c|c|c||c|c|c|c|c|}
\hline
$\abs{m_{3/2}}$ & $\abs{\mu^{'}}$ & $\theta_{\CP}$ & $f_{wo}$ & ${\rm T_{RH}}$ & $\beta$ & $\frac{Y_{B}^{\ast}}{Y_{B}^{\rm (obs)}}$ & $\frac{m_{\nu}}{m_{\nu}^{\rm (obs)}}$  \\ 
\hline \hline
$10^{8}$ & $9 \cdot 10^{7}$ & $0.1$ & $0.05$ & $10^{16}$ & $1\cdot10^{-4}$ & $0.4$ & $3\cdot 10^{4}$ \\
$10^{10}$ & $3\cdot10^{9}$ & $1$ & $1.0$ & $10^{16}$ & $7\cdot10^{-4}$ & $3$ & $40$ \\
$10^{12}$ & $3\cdot 10^{11}$ & $0.5$ & $0.7$ & $10^{16}$ & $3\cdot10^{-4}$ & $1$ & $0.4$ \\
$10^{13}$ & $10^{13}$ & $0.1$ & $0.05$ & $10^{15}$ & $2\cdot10^{-4}$ & $0.1$ & $0.4$ \\
$10^{14}$ & $9\cdot10^{13}$ & $0.1$ & $0.1$ & $10^{16}$ & $2\cdot10^{-4}$ & $1$ & $4\cdot 10^{-2}$ \\
\hline
\end{tabular}
\end{center}
\caption{\label{tab:HL}Typical parameter sets for the model with $W_{\LV} = \hat{H}_u \hat{L}$ that produce the observed baryon asymmetry. The input parameters are the gravitino mass ($\abs{m_{3/2}}$), the mixing mass scale ($\mu^{\prime} \sim \sqrt{B^u} \sim \sqrt{B^d}$), the CP-violating phase ($\theta_{\CP}$) in radians, the washout factor ($f_{wo}$), and the reheat temperature ($T_{RH}$). Dimensionful parameters are expressed in units of $\GeV$.  We estimate the baron asymmetry, $Y_B^{\ast}$ using \eref{eq:YB_estimate} along with the approximate estimate for $\beta$ given by \eref{eq:HL_beta_estimate}, and the ratio of the neutrino mass $m_{\nu}$ to the observed value, $m_{\nu}^{\rm (obs)}\approx1 \eV$, by \eref{eq:HL_mnu}, setting $m_{\tilde{X}}\sim m_{3/2}$ and $t_{\beta}=1$. 
}
\end{table}%

%===========
Finally, gravitino leptogenesis requires a firm lower bound on the reheat temperature $T_{RH}^{\rm min} \approx 10^{12} \GeV$ in order to generate a sufficiently large population of gravitinos to account for the observed baryon asymmetry in their decays.  
At present, constraints on the CMB tensor-to-scalar ratio, $r \propto T_{RH}^{1/4}$, give an upper bound on the reheat temperature, $T_{RH}^{\rm max} \approx 10^{16} \GeV$, so that significant improvements in sensitivity beyond those likely in the near future would be required to use this limit to probe this scenario.

%----------------------------------------------------------------
% Acknowledgments
%----------------------------------------------------------------
\begin{acknowledgments}
This work was supported by the DOE under Grant No.\ DE-SC0008016, and also by ANU. 
\end{acknowledgments}

%===========================
% Appendix
%===========================
\begin{appendix}

%----------------------------------------------------------------
% Notation
%----------------------------------------------------------------
\section{Notation}\label{app:Notation}

%===========
\begin{table}[t]
\begin{center}
\begin{tabular}{|l|c|c|c|c|c|c|}
\hline
Field & $\SU{3}_C$ & $\SU{2}_L$ & $\U{1}_Y$ & $\U{1}_R$ & $\U{1}_B$ & $\U{1}_L$ \\
\hline
$\hat{Q}_i$ & $\mathbf{3}$ & $\mathbf{2}$ & $1/6$ & $1$ & $1/3$ & $0$ \\
$\hat{U}^c_i$ & $\bar{\mathbf{3}}$ & $\mathbf{1}$ & $-2/3$ & $-1$ & $-1/3$ & $0$ \\
$\hat{D}^c_i$ & $\bar{\mathbf{3}}$ & $\mathbf{1}$ & $1/3$ & $-1$ & $-1/3$ & $0$ \\
$\hat{L}_i$ & $\mathbf{1}$ & $\mathbf{2}$ & $-1/2$ & $-1$ & $0$ & $1$ \\
$\hat{E}^c_i$ & $\mathbf{1}$ & $\mathbf{1}$ & $1$ & $1$ & $0$ & $-1$ \\
$\hat{H}_u$ & $\mathbf{1}$ & $\mathbf{2}$ & $1/2$ & $0$ & $0$ & $0$ \\
$\hat{H}_d$ & $\mathbf{1}$ & $\mathbf{2}$ & $-1/2$ & $0$ & $0$ & $0$ \\
\hline
$\hat{G}$ & $\mathbf{1}$ & $\mathbf{1}$ & $0$ & $0$ & $0$ & $0$ \\
$\hat{g}$ & $\mathbf{8}$ & $\mathbf{1}$ & $0$ & $0$ & $0$ & $0$ \\
$\hat{W}$ & $\mathbf{1}$ & $\mathbf{3}$ & $0$ & $0$ & $0$ & $0$ \\
$\hat{B}$ & $\mathbf{1}$ & $\mathbf{1}$ & $0$ & $0$ & $0$ & $0$ \\
\hline
\end{tabular}
\end{center}
\caption{\label{tab:fields}
The MSSM field content and charges.  
}
\end{table}%

%===========
We include here a summary of the notation and conventions used in this paper.  
The field content is summarized in \tref{tab:fields}.  
The R-charge is given by $Q_R = 3(B-L)$.  
The quark and lepton superfields may be expanded in flavor space as 
\begin{align}
\begin{array}{l}
	\hat{Q}_i = \bigl\{ \hat{Q}_1 \, , \, \hat{Q}_2 \, , \, \hat{Q}_3 \bigr\} \\
	\hat{U}^c_i = \bigl\{ \hat{U}^c \, , \, \hat{C}^c \, , \, \hat{T}^c \bigr\} \\
	\hat{D}^c_i = \bigl\{ \hat{D}^c \, , \, \hat{S}^c \, , \, \hat{B}^c \bigr\} \\
	\hat{L}_i = \bigl\{ \hat{L}_1 \, , \, \hat{L}_2 \, , \, \hat{L}_3 \bigr\} \\
	\hat{E}^c_i = \bigl\{ \hat{e}^c \, , \, \hat{\mu}^c \, , \, \hat{\tau}^c \bigr\} \\
\end{array} \per
\end{align}
The left-chiral superfields may be expanded in superspace as 
\begin{align}\label{eq:expandsuperspace}
\begin{array}{lclcl}
	\hat{Q}_1 = \bigl( \tilde{q}_1 \, , \, q_1 \bigr) & \qquad & 
	\hat{Q}_2= \bigl( \tilde{q}_2 \, , \, q_2 \bigr) & \qquad & 
	\hat{Q}_3 = \bigl( \tilde{q}_3 \, , \, q_3 \bigr) \\
	\hat{U}^c = \bigl( \tilde{u}^c \, , \, u^c \bigr) & \qquad &
	\hat{C}^c = \bigl( \tilde{c}^c \, , \, c^c \bigr) & \qquad & 
	\hat{T}^c = \bigl( \tilde{t}^c \, , \, t^c \bigr) \\
	\hat{D}^c = \bigl( \tilde{d}^c \, , \, d^c \bigr) & \qquad &
	\hat{S}^c = \bigl( \tilde{s}^c \, , \, s^c \bigr) & \qquad &
	\hat{B}^c = \bigl( \tilde{b}^c \, , \, b^c \bigr) \\
	\hat{L}_1 = \bigl( \tilde{\ell}_1 \, , \, \ell_1 \bigr) & \qquad &
	\hat{L}_2 = \bigl( \tilde{\ell}_2 \, , \, \ell_2 \bigr) & \qquad &
	\hat{L}_3 = \bigl( \tilde{\ell}_3 \, , \, \ell_3 \bigr) \\
	\hat{e}^c = \bigl( \tilde{e}^c \, , \, e^c \bigr) & \qquad &
	\hat{\mu}^c = \bigl( \tilde{\mu}^c \, , \, \mu^c \bigr) & \qquad &
	\hat{\tau}^c = \bigl( \tilde{\tau}^c \, , \, \tau^c \bigr) 
\end{array} \\
{\rm and}Œ
\hspace{1cm}
\begin{array}{lclcl}
	\hat{H}_u = \bigl( h_u \, , \, \tilde{h}_u \bigr) & \qquad & & \qquad &  \\
	\hat{H}_d = \bigl( h_d \, , \, \tilde{h}_d \bigr) & \qquad & & \qquad &  \\
\end{array}
\end{align}
where the first entry is a complex scalar scalar field and the second is a left-chiral Weyl spinor.  
Since we are interested in gravitino decays prior to electroweak symmetry breaking, it is convenient to use the two-component spinor notation.  
In this formalism $\chi_{\alpha}$ transforms in the $(1/2, 0)$ representation of the Lorentz group while $\chi^{\dagger}_{\dot{\alpha}}$ transforms in the $(0,1/2)$ representation (see \cite{Dreiner:2008tw} for a review).  
Each of the spinors in \eref{eq:expandsuperspace} is left-chiral, including those denoted with a ``c'' superscript.  
The three vector superfields are given by 
\begin{align}
\begin{array}{l}
	\hat{g} = ( g \, , \, \tilde{g}) \\
	\hat{W} = ( W \, , \, \tilde{W}) \\
	\hat{B} = ( B \, , \, \tilde{B}) 
\end{array} \per
\end{align}
Collectively, we will use $\tilde{X}$ to denote the gauginos $\tilde{g}$, $\tilde{W}$, and $\tilde{B}$.  
Finally the gravitino is denoted as $\tilde{G}$.  
In writing particle reactions, the CP conjugate particle is denoted by a bar.  
For example, the spinor field $e^c$ has quanta, denoted as $e^c$ and $\bar{e^c}$ in particle reactions, with opposite charge and chirality.  

%----------------------------------------------------------------
% Gravitino Baryogenesis Details
%----------------------------------------------------------------
\section{Gravitino Baryogenesis Details}\label{app:GravitinoBaryogenesis}

%===========
In this appendix we provide additional details of the gravitino decay calculation of \rref{Cline:1990bw} that are relevant for our analysis.  
The baryon asymmetry parameter is given by \eref{eq:UDD_beta_def}, which is reproduced here for convenience:  
\begin{align}\label{eq:UDD_beta_appendix}
	\beta = \sum_{q, \tilde{q}} \Biggl\{ &
	\left( {\rm BR}[ \tilde{G} \to q \bt{q}] - {\rm BR}[ \tilde{G} \to \bar{q} \tilde{q}] \right) \nn
	&+ \sum_{\tilde{X} = \tilde{g}, \tilde{Z}, \tilde{\gamma}} {\rm BR} [ \tilde{G} \to X \tilde{X} ] 
	\left( {\rm BR}[ \tilde{X} \to q \bt{q}] - {\rm BR}[ \tilde{X} \to \bar{q} \tilde{q}] \right)
	\Biggr\} 
	{\rm BR}[ \bt{q} \to qq ]
	\per
\end{align}
In principle the sum runs over all quark and squark species, but in light of the interactions in \eref{eq:UDD_LBV}, CP violation is only carried by quanta of the fields $s^c, b^c, t^c, \tilde{s}^c, \tilde{b}^c,$ and $\tilde{t}^c$.  
Since these fields carry B-number of $-1/3$ (see \aref{app:Notation}), the sum is over $q = \overline{s^c}, \overline{b^c}, \overline{t^c}$ and $\tilde{q} = \bt{s^c}, \bt{b^c}, \bt{t^c}$.  
The expression \eref{eq:UDD_beta_appendix} assumes that ${\rm BR}[ \bt{q} \to qq ] = {\rm BR}[ \tilde{q} \to \bar{q}\bar{q} ] \equiv {\rm BR}_{\BV}$ since any difference must be proportional to the CP violating parameter, which would yield a higher order correction to $\beta$.  

%===========
These various branching ratios in \eref{eq:UDD_beta_appendix} can be calculated exactly \cite{Cline:1990bw}, but since we are primarily interested in exploring the possibility of obtaining the correct order of magnitude for the resulting baryon asymmetry here, we will assume the hierarchy of mass scales $m_{3/2} > m_{\tilde{X}} > m_{\tilde{q}} > m_q$.  
In this limit, the branching ratios can be estimated from dimensional analysis up to undetermined $O(1)$ prefactors.  
Since the gravitino has a universal gravitational strength coupling, it decays with equal probability into every light species.  
Then, the branching fraction into the vector supermultiplets can be estimated as 
\begin{align}
	{\rm BR}[ \G \to X \X ] \approx \frac{\mathcal{C}_{\X}}{N_{\rm eff}}
\end{align}
where $\mathcal{C}_{\X}$ is the dimension of the adjoint representation of the gauge group corresponding to the gaugino $\X$, i.e., 
\begin{align}
	\mathcal{C}_{\tilde{B}} = \mathcal{C}_{\tilde{Z}} = \mathcal{C}_{\tilde{\gamma}} = 1 
	\qquad , \qquad 
	\mathcal{C}_{\tilde{W}} = 3 
	\qquad , \qquad 
	\mathcal{C}_{\tilde{g}} = 8 \com
\end{align}
and $N_{\rm eff} \simeq 16$ is given by \eref{eq:Neff_def}.  

%===========
The differential branching fractions into the quark -- squark final states are nonzero due to an interference between graphs of the form shown in \fref{fig:UDD_CPV}.  
The CP violation arises from the relative phase of the coupling $A_{332}^{\prime \prime}$ and the gravitino or gaugino mass parameter.  
Up to factors of $O(1)$, the interference can be estimated as 
\begin{align}\label{eq:UDD_diferential_BR}
	\left( {\rm BR}[ \tilde{V} \to q \bt{q}] - {\rm BR}[ \tilde{V} \to \bar{q} \tilde{q}] \right) \sim \alpha_{332}^{\prime \prime} \frac{{\rm Im}\left[ A_{332}^{\prime \prime} m_{\tilde{V}} \right]}{\abs{m_{\tilde{V}}}^2} \times \begin{cases} 1 & \tilde{V} = \tilde{g} , \tilde{Z} , \tilde{\gamma} \\ \frac{1}{N_{\rm eff}} & \tilde{V} = \G  \end{cases}
\end{align}
where $\alpha_{332}^{\prime \prime} \equiv \abs{\lambda_{332}^{\prime \prime}}^2 / 4 \pi$.  
For simplicity one can assume that there is a universal CP-violating parameter $\theta_{\CP} = {\rm Arg}[A_{332}^{\prime \prime} m_{\tilde{V}}]$ for both the gravitino and gaugino masses, and therefore 
\begin{align}
	\frac{{\rm Im}[A_{332}^{\prime \prime} m_{\tilde{V}}]}{\abs{m_{\tilde{V}}}^2} = \frac{\abs{A_{332}^{\prime \prime}}}{\abs{m_{\tilde{V}}}}  \sin \theta_{\CP} \per
\end{align}
One can also assume that there is a universal gaugino mass $m_{\tilde{X}} = m_{\tilde{\gamma}} = m_{\tilde{Z}} = m_{\tilde{g}}$.  
In this case to evaluate \eref{eq:UDD_beta_appendix} one need only sum the possible decay channels.  
The gravitino, photino, and zino each have $9$ decay channels, given by summing over the combinations of three flavors ($s^c, t^c, b^c$) and three colors.  
Since the gluino is colored, it has only $3$ channels, given by the sum over flavors alone.  
After these simplifications, \eref{eq:UDD_beta_appendix} becomes
\begin{align}
	\beta \sim \alpha_{332}^{\prime \prime} \sin \theta_{\CP} \frac{1}{N_{\rm eff}} {\rm Max} \left[ 9 \frac{\abs{A_{332}^{\prime \prime}}}{\abs{m_{3/2}}} \, , \, 42 \frac{\abs{A_{332}^{\prime \prime}}}{\abs{m_{\tilde{X}}}} \right] {\rm BR}_{\BV} \per
\end{align}
%Gravitino = 9 = 3(flavor) * 3(color)
%Photino = Zino = 9 = 3(flavor) * 3(color)
%Gluino = 3(flavor)
%Gaugino Total = 42 = 1(photino)*9 + 1(zino)*9 + 8(gluino)*3
Since the relative $O(1)$ factors between the gravitino and gaugino contributions have been left unspecified in \eref{eq:UDD_diferential_BR} one cannot precisely sum the two contributions, and instead one can simply estimate $\beta$ by taking the larger of the two.  
The numerical factors of $9$ and $42=9+9+8\cdot 3$ arise from counting the decay channels.  
Making the further assumption $m_{\tilde{X}} \approx m_{3/2}$ one obtains \eref{eq:UDD_beta_numerical}

%===========
One may worry that strong constraints on the RPV coupling $\lambda_{332}^{\prime \prime}$ would cause ${\rm BR}_{\BV}$, and therefore $\beta$, to be too small for baryogenesis to succeed.  
However, if alternate decay channels are kinematically blocked by spectral constraints then the branching fraction can be $O(1)$.  
For instance, if the squark is the LSP then ${\rm BR}_{\BV} = 1$.  
More generally, one can suppose that the gluinos are light and the other gauginos are heavy, and in this case the branching fraction is estimated as 
\begin{align}
	\eBV \approx 1 - O \left( \alpha_s / \alpha_{332} \right) \per
\end{align}
This helps to evade strong constraints on B-number violation.

\end{appendix}

%----------------------------------------------------------------
% References
%----------------------------------------------------------------
\bibliographystyle{JHEP}
\bibliography{refs--GravLeptogen}

\end{document}